\documentclass[fleqn,usenatbib]{mnras}

\usepackage{newtxtext,newtxmath}

\usepackage[T1]{fontenc}

\DeclareRobustCommand{\VAN}[3]{#2}
\let\VANthebibliography\thebibliography
\def\thebibliography{\DeclareRobustCommand{\VAN}[3]{##3}\VANthebibliography}

\usepackage{graphicx}	
\usepackage{amsmath}	
\usepackage{xcolor}
\usepackage{ulem}

\def\msun{\,{\rm M_\odot}}

\title[cosmic BBHs]{Unveiling  
early black hole growth with multi-frequency\\ gravitational wave observations
}
\author[Valiante et al.]{
Rosa Valiante,$^{1,2}$\thanks{E-mail: rosa.valiante@inaf.it}
Monica Colpi,$^{3,4}$
Raffaella Schneider$^{5,1,2}$, Alberto Mangiagli$^{3,4}$,
\newauthor
Matteo Bonetti$^{3}$, Giulia Cerini$^{6}$,
Stephen Fairhurst$^{7}$, Francesco Haardt$^{8,4}$, Cameron Mills$^{7}$
\newauthor
and Alberto Sesana$^{3}$
\\
$^{1}$ INAF-Osservatorio Astronomico di Roma, via di Frascati 33, I-00078 Monteporzio Catone, Italy\\
$^{2}$ INFN, Sezione di Roma I, P.le Aldo Moro 2, I-00185 Roma, Italy\\
$^{3}$ Department of Physics,  University of Milano - Bicocca,  Piazza della Scienza 3,I20126 Milano, Italy\\
$^{4}$ National Institute of Nuclear Physics INFN, Milano - Bicocca, Piazza della Scienza 3, 20126 Milano, Italy\\
$^{5}$ Dipartimento di Fisica, Universitá di Roma `La Sapienza', P.le Aldo Moro 2, I-00185 Roma, Italy\\
$^{6}$ Department of Physics, University of Miami, 1320 Campo Sano Drive, Coral Gables, FL 33124, USA \\
$^{7}$ Gravity Exploration Institute, School of Physics and Astronomy, Cardiff University, Cardiff, UK, CF24 3AA, UK\\
$^{8}$ DiSAT, Università dell'Insubria, via Valleggio 11, 22100 Como, Italy  
}

\date{Accepted XXX. Received YYY; in original form ZZZ}

\pubyear{2020}

\begin{document}
\label{firstpage}
\pagerange{\pageref{firstpage}--\pageref{lastpage}}
\maketitle

\begin{abstract}
{
Third Generation ground based Gravitational Wave Interferometers, like the Einstein Telescope (ET),  Cosmic Explorer (CE), and the Laser Interferometer Space Antenna (LISA) will detect coalescing binary black holes over a wide mass spectrum and across all cosmic epochs. We track the cosmological growth of the earliest light and heavy seeds that swiftly transit into the supermassive domain using a semi-analytical model for the formation of quasars at $z = 6.4,2$ and $0.2$, in which we follow black hole coalescences driven by triple interactions. We find that light seed binaries of several $10^2 \msun$ are accessible to ET with a signal-to-noise ratio (S/N) of $10-20$ at $6<z<15$. They then enter the LISA domain with larger S/N as they grow to a few $10^4 \msun$. Detecting their gravitational signal would provide first time evidence that light seeds form, grow and dynamically pair during galaxy mergers. The electromagnetic emission of accreting black holes of similar mass and redshift is too faint to be detected even for the deepest future facilities. ET will be our only chance  to discover light seeds forming at cosmic dawn. At $2<z<8$, we predict a population of "starved binaries", long-lived marginally-growing light-seed pairs, to be loud sources in the ET bandwidth (S/N$>20$). Mergers involving heavy seeds ($\sim 10^5\msun -10^6 \msun$) would be within reach up to $z=20$ in the LISA frequency domain. {The lower-$z$ model predicts $11.25 \, (18.7)$ ET (LISA) events per year, overall}. 
}
\end{abstract}



\section{Introduction}
The discovery of luminous quasars powered by accretion onto $10^9\msun -10^{10} \msun$ supermassive black holes (SMBHs) at redshift as early as $z\sim 7.5$ \citep{Banados18, Yang2020}, only $\sim 800$ Myr after the Big Bang, has revolutionized our view on how these giants formed before the epoch of cosmic reionization \citep{Banados16}. They represent the tip of an underlying population of much fainter Active Galactic Nuclei \citep [AGN,] []{Matsuoka18} that are the least known in terms of basic demographics, birth, and growth. As gas is likely the primary fuel for their growth \citep{Marconi04,Merloni04b, Kormendy13, Trakhtenbrot2020review}, this observation hints to the existence, at redshifts $z>7$, of a population of {\it seed} black holes (BHs) of yet unconstrained initial mass, in the range from about $\sim 100 \msun$  to $\sim 10^5 \msun$ from  
which the giants have grown. This interval is often referred to as intermediate (between stellar-mass BHs and SMBHs), with {\it light seeds} in the range between $\sim 10^{2}\msun$ and a few $10^3\msun$, and {\it heavy seeds} in the range between $10^4 \msun$ and $10^6 \msun$ as extremes \citep{Valiante17review}.

The origin of seeds is not known yet, nor the mechanisms leading to their swift evolution to become  high$-z$ quasars \citep{Volonteri10,Schleicher13,Latif16,Johnson16, Inayoshi2019}. 
Various avenues of formation have been proposed:
\newline
\newline
\noindent
{\bf Light seeds:} massive stars collapsing into stellar BHs beyond the pair instability gap \citep{HegerWoosley10}, with masses of a few $10^2\msun$  forming in metal-free/poor dark matter (DM) halos at redshifts $z$ as large as $\sim 20 - 30$ \citep{Abel02, Heger03, MadauRees01, Yoshida08, Hirano14, Hirano15};\\
{\bf Medium-weight seeds:}  very massive stars, resulting from runaway stellar mergers, in compact star  clusters forming at $z \sim 10$ \citep{Devecchi12, Mapelli16, Reinoso18}. 
Here stellar masses of $\sim 200-10^3\msun$ are not set by the fragmentation properties of the birth gas clouds but by stellar collisions ruled by the dynamics inside the earliest dense nuclear star clusters. Alternatively, they may form in runway gravitational wave (GW) driven coalescences of stellar BHs in star clusters subject to major gas inflows, at the centre of pre-galactic discs forming at  $z\sim 10$ \citep{Davies11,Lupi14};\\
{\bf Heavy seeds:} supermassive (proto)-stars  of $\sim 10^{4-6}\msun$ growing through continued and fast accretion within their birth clouds, collapsing directly onto a BH, the so-called direct collapse BH (DCBH) scenario, driven by general relativistic instabilities or fuel exhaustion \citep{BrommLoeb04, Begelman06, IO12, Inayoshi14, Umeda16}. These are considered to be rare seeds due to their contrived birth environmental conditions \citep{Agarwal12, Latif13,Dijkstra14,Habouzit16, Chon16, Valiante16, Regan17}. 
Intense UV radiation from adjacent star forming regions and large infall rates of metal-free/poor gas are required to suppress fragmentation of the birth cloud and to feed the central proto-star. Even in slightly enriched halos ($Z<10^{-3} \, \rm Z_\odot$), where fragmentation takes place, infalling, metal-poor, material preferentially feeds the primary proto-star (the first to form in the cloud) that grows super-massive \citep[the so-called super-competitive accretion scenario][]{Chon20}. 
Alternatively, the formation of heavy seeds may be aided by dynamical heating during rapid mass growth of low-mass halos in over-dense regions at high redshifts \citep{Wise19}, or by massive nuclear inflows in major gas-rich galaxy mergers at lower redshift \citep{Mayer15}.

Currently, the only way to infer information on BHs of $\sim 10^5 \msun$ is by looking at local dwarf galaxies \citep{Reines15, Baldassare15, Mezcua2016, Mezcua2018} where observational signatures of seed formation are expected to be strong \citep{Habouzit16}. 
Although the faint-end tail of the $z\sim 6$ AGN luminosity function has been sampled down to absolute magnitude of $M_{1459}=-22$ mag \citep{Matsuoka18}, 
no observational signatures of fainter AGN, possibly powered by BHs of $< 10^7 \msun$, have been found at higher redshifts. The non-detection of faint high-z AGN may be a consequence of their low active fraction \citep[$\sim 0.1\%$ at $z>7$][]{Pezzulli17} and/or of their relatively low number density \citep[][but see \citealt{Wise19}]{Habouzit16, Valiante16,Cowie20}. 

In the next decades, with the advent of the foremost electromagnetic (EM) facilities and of the next generation of ground- and space-based gravitational wave (GW) interferometers, breakthrough in this field will be accomplished exploiting jointly the power of traditional Astrophysics with the nascent multi-frequency GW Astronomy. 

{\it Light waves} on the one side:
the Square Kilometer Array in radio, the James Webb Space Telescope ({\it JWST}) and the Extremely Large Telescope in the optical and near-infrared, the Advanced Telescopes for High Energy Astrophysics {\it Athena} and the mission-concept {\it Lynx} in the X-rays,
will provide new information on
the earliest accreting BHs, the dimmest AGN  of the low-mass tail of SMBH population, binary or$/$and multiple AGN in interacting systems, and ultimately will let us identify the EM counterparts of the loudest GW signals from merging massive BHs \citep{Dalcanton2019,SesanaNature2020}. 

{\it Gravitational waves} on the other side: third generation ground-based interferometers such as Einstein Telescope \citep[ET,][]{ET10,ET12} and  Cosmic Explorer \citep[CE][]{CE17, Reitze2019} will capture the GW signal from millions of coalescing stellar binary BHs (BBHs) detectable out to $z\sim 10-15$. In particular ET, with a higher sensitivity al the lowest frequencies around 3-10 Hz has the potential of discovering mergers of BBHs with masses up to a few $100\msun$ characteristic of the earliest stellar and seed BH populations and BBHs of a few $10^3\msun$ at moderate redshifts \citep{Kalogera19,Maggiore19}.
Space-based interferometers such as the Laser Interferometer Space Antenna (LISA), the interferometer 
TianQin under design \citep{Luo2016}  and the proposed Taiji program \citep{Ruan2018} will instead detect the GW signals from massive BBH coalescences (from $\sim 10^4 \msun$ up to a about  $\sim 10^7 \msun$) across all cosmic ages providing the first ever census of this new population of BHs that formed in the aftermath of galaxy collisions   \citep{LISA17,Colpi19}.
{
Thus, future GW observatories together will detect the signal emitted by coalescing binary BHs over a wide mass spectrum, from the stellar to the massive, through the formation of seeds, and across all cosmic epochs.
}

Seeds are expected to grow via accretion of surrounding gas in primeval DM halos. Their growth might be Eddington limited leading to an $e$-fold increase in the mass on timescales of a few
$100$ Myr if uninterrupted.
Growth may occur at super-Eddington rates if seeds are surrounded by radiatively inefficient slim discs \citep{2014ApJ...784L..38M,2015ApJ...804..148V,Pezzulli16},
or at supra-exponential rates if embedded in star 
clusters fed by dense cold gas, expected to be ubiquitous in the high redshift Universe \citep{Alexander2014}. But,  BHs invariably participate in the assembly of cosmic structures during their evolution, possibly growing also through coalescences, in addition to gas accretion \citep{Volonteri03, Sesana2007, Valiante16}.
This implies that seed BHs might pair and merge shortly after 
their formation in the earliest halo-halo merger events, becoming high-$z$ sources of GWs at frequencies of $\sim 3$ - $10$ Hz, in the ET
frequency band (light seeds), and/or $100\mu$Hz - 100 mHz, the LISA domain (medium-weight and heavy seeds).

In this paper we aim at exploring the emergence of {\it cosmologically-driven} pairs of seed BHs merging in the aftermath of halo-halo collisions, following their growth via accretion and mergers to track their swift transit across the ET and LISA bandwidths, as GW sources. To this purpose we improve upon \textsc{GAMETE/QSOdust (GQd)}, the Semi-Analytical Model (SAM) presented in  \citet{Valiante16, Valiante18a}. 
Developed to model the formation and evolution of high-$z$ quasars, \textsc{GQd} includes a refined seeding prescription for both light and heavy seeds combing chemical and radiative properties of the environment in halos selected among $z>10$ progenitors of $z>6$ quasars.

In addition, in  \citet[][]{Valiante18a, Valiante18b} we followed the early growth of a seed via gas accretion only inside  an evolving unperturbed halo, before the information on its birth environment (and hence on the nature of the BH seed) was erased as a consequence of a halo-halo merger. By processing the radiation emitted by the stars and accreting BHs through gas and dust, we showed that the most massive ($>10^6 \msun$) and rapidly growing seeds would be easily detected by future (EM) missions, like 
{\it Athena} and {\it JWST} \citep[][]{Pacucci15, Natarajan17}, up to $z\sim 15$ \citep{Valiante18b}.  By contrast, lighter accreting BHs  with mass $\lesssim {10^3} \msun$ would remain undetectable due to their weaker emission, showing the limiting power of EM observations in detecting seed BHs.  In this paper we aim at exploring  whether future GWs telescopes would allow us to discover in a unique way  the formation and evolution of the earliest seeds and their potential link with SMBHs \citep{Colpi2019-book}.

Using \textsc{GQd}, we focus here on the histories of three DM halos, of equal mass, each hosting a quasar shining at a different redshift: $z_{\rm QSO}\sim 6.4$, near the epoch of reionization of the intergalactic hydrogen, at $z_{\rm QSO}\sim 2$, near the peak of the cosmic SFR density in the Universe, and at $z_{\rm QSO}=0.2$, during the fading of the AGN activity and quenching of the SFR.
We follow the hierarchical formation pathways of these quasars by describing seed growth ruled by accretion episodes and mergers in multiple DM halo collisions, including in \textsc{GQd} a prescription to track their dynamics down to coalescence, driven by triple BH interactions \citep{Bonetti16,Bonetti18b}.

The paper is organized as follows. The semi-analytical approach is summarized in Section~\ref{qso-evolution} while in Section~\ref{sec:BBHs} the new features of the model are described. In Sections~\ref{sec:QSOs} and \ref{sec:EM} the emergence of binary black holes within our model is analyzed in view of the future GW and EM facilities. A critical discussion of our approach is presented in Section~\ref{sec:discussion}. Finally, our main conclusions are drawn in Section~\ref{sec:conclusions}.

\section{The quasar evolution model}\label{qso-evolution}
In this section we summarize the main features of our data-constrained SAM, \textsc{GQd}, and defer the interested reader to \citet{Valiante16, Valiante18a, Valiante18b}, and references therein, for details.
The model follows the formation and evolution of 
individual quasars, powered by accretion onto supermassive black holes (SMBHs), and their host galaxies, observed at high redshift, with particular attention to $z>6$ systems, like SDSS J1148+5251 (J1148) at $\rm z_{\rm QSO}=6.4$, \citep{Valiante11, Valiante16}. 
\textsc{GQd} has been extensively tested against a sample of $z_{\rm QSO}>5$ quasars, well reproducing their observed properties \citep{Valiante14}. For the purposes of the present work, we extend the analysis to lower redshift analogs, i.e. quasars at $z_{\rm QSO}=2$ and $z_{\rm QSO}=0.2,$ respectively. 
The evolution of each DM halo is described
using  semi-analytically reconstructed merger histories.

\subsection{Dark Matter Halo}\label{sec:DMhalo}

With \textsc{GQd}, we produce for each simulated quasar ten merger tree realizations of a DM halo of  $M_0=10^{13}\rm{M_\odot}$, in which the luminous quasar is expected to reside.\footnote{It is commonly believed that $[10^{12} - 10^{13}]\msun$ host DM halos are required to match the observed space density of $z \sim 6$ quasars (\citealt{Fan04} and see \citealt{Valiante11} for a discussion.} This DM halo is decomposed into progressively less massive fragments, called progenitors, through a binary Monte Carlo algorithm with mass accretion based on the Extended Press-Schechter formalism \citep{PS74}.

At a given redshift $z$ along the merger tree, the minimum mass of a resolved structure (virialized progenitor), i.e. the merger tree mass resolution, is described as
\begin{equation}
    M_{\rm res}(z)=10^{-3}M_0\biggl(\frac{1+z}{1+z_{\rm QSO}}\biggr)^\beta, 
    \numberwithin{equation}{section}
\end{equation}\label{eq:resolutionMass}
where $M_0 = 10^{13}$ M$_\odot$ is the same for the three quasars and the parameter $\beta$ is assumed to be $-7.5$, $-4.3$ and $-3.0$, for $z_{\rm QSO}=6.4$, $2$, and $0.2$, respectively \citep{Valiante16}, so that at $z=24$ ($z=z_{\rm QSO}$) $M_{\rm res}\sim 10^{6} \, (10^{10}) \msun$. 
Non resolved structures with $M<M_{\rm res}$ account for the external, intergalactic medium (IGM) from which progenitor halos accrete mass.

The characteristic redshift interval of the merger tree models, 
$\Delta z$,  the functional form of the mass resolution and the value of the parameter $\beta$ has been chosen to (i) resolve mini-halos (i.e. those DM progenitors with virial temperatures in the range $1200$ K $\leq T_{\rm vir}<10^4$ K) at high redshift, (ii) prevent the formation of multiple fragments ($>2$ per progenitor halo, as required by the binary algorithm), (iii) reproduce the Extended Press-Schechter halo mass functions and (iv) limit the computational times. 
These requirements determine the redshift distribution and total number of progenitors forming between $z=24$ and $z_{\rm QSO}$, which is higher for lower $z_{\rm QSO}$ simulations.

According to Eq.~\ref{eq:resolutionMass} mini-halos of $\sim 10^6 - 10^8 \msun$ are resolved 
at $z > 13$, $8$ and $5$ in the merger trees of the 
$z_{\rm QS0}=6.4$, $2$ and $0.2$ simulated quasar hosts, respectively.
These low-mass halos are expected to be the first formation sites of Population~III stars, at $z\sim 20-30$, and of light seeds. Along each reconstructed merger trees, \textsc{GQd} consistently follows the evolution of each progenitor galaxy and its nuclear BH, running forward in time from $z=24$ to $z_{\rm QSO}$.

{
The adopted resolution mass does not have a significant impact on the analysis presented here since, close to the final redshift, accretion and merging of low-mass halos increase their mass above the resolution. Furthermore, chemical and radiative feedback inhibit the formation of black hole seeds when $z < 17 \, (13, 12)$ for the quasar models with $z_{\rm QSO}=6.4 \, (2, 0.2)$  (see Section~\ref{sec:seedsToBBHs}).
}

\subsection{Quasar's progenitor galaxies}
The (co-)evolution of BHs and their host galaxies is a complex process, regulated by the interplay between chemical, mechanical and radiative feedback. In the framework of mainstream structure formation scenarios, seeds grow by accreting at a rate regulated by the reservoir of dense, cold gas present in their neighbourhood. This, in turn, is set by the baryon cycle of the forming host galaxy that gains mass through gas inflows from the external IGM, consumes mass to fuel star formation, and loses mass via winds powered by supernova explosions and by the radiation that the BH feeds back into the interstellar medium (ISM).

Mass exchanges with the IGM, genetic (in-situ) ISM metal enrichment of the galaxies and the intensity of the permeating UV field all contribute to determine the efficiency of star formation (especially in mini-halos), the duration of the Pop~III star forming epoch and the number and nature of BH seeds that form.

\subsubsection{Star formation}
In each progenitor galaxy, we convert gas into stars at a rate that is given by:
\begin{equation}
{\rm SFR} = f_{\rm cool} \, M_{\rm gas}\, \epsilon/t_{\rm dyn}(z),
\numberwithin{equation}{section}
\label{eq:sfr}
\end{equation}
where SFR is the star formation rate and $t_{\rm dyn}(z)=R_{\rm vir}/v_{e}$ is the redshift-dependent dynamical timescale (being $R_{\rm vir}$ and $v_{e}$ the halo virial radius and escape velocity).
In our model stars form through a series of quiescent ($\epsilon = \epsilon_{\rm quiesc}$) and major-merger enhanced bursts ($\epsilon = \epsilon_{\rm quiesc}+\epsilon_{\rm burst}$). 
The quiescent star formation efficiency is a free parameter of the model and the choice of its value is discussed in Section \ref{sec:QSOs}. 
The parameter $\epsilon_{ \rm burst}$ accounts for the efficiency enhancement due to major galaxy mergers, that is the coalescences of two DM halos with mass ratios, $\mu_{\rm DM} > 1/4$ (least massive over most massive). 
In \textsc{GQd} $\epsilon_{\rm burst}$ is a function of $\mu_{\rm{DM}}$, computed as a Gaussian distribution with $\sigma_{\rm burst}=0.05$ (we have $\epsilon_{\rm burst}=8$ for $\mu_{\rm DM}=1/4$; see \citealt{Valiante11}).

Finally, the quantity $f_{\rm cool}$ is the ratio between the total mass of gas enclosed in the halo virial radius and the gas mass within the "cooling radius" $r_{\rm cool}$, the radius at which the cooling time, $t_{\rm cool}$ equals the free fall time, $t_{\rm ff}$.
The value of $f_{\rm cool}$ represents the reduced star formation efficiency of mini-halos ($f_{\rm cool}<1$) with respect to atomic cooling halos ($T_{\rm vir}\geq 10^4$ K, $f_{\rm cool}=1$), as described in \citet[][]{Valiante16} and \citet[][]{deBennassuti17}. In mini-halos, in fact, the fraction of the available gas that can cool and form stars strongly depends on halo properties (virial temperature, redshift and gas metallicity) and on the intensity of illuminating far UV radiation, that can photo-dissociate $\rm H_2$ molecules, the main coolant in these halos.

For each stellar population formed via Eq. \ref{eq:sfr} we adopt a Larson initial mass function \citep[IMF][]{Larson98} to describe the stellar mass spectrum. The first generation of stars (Pop~III stars) forms in pristine/metal poor galaxies with metallicity $Z < Z_{\rm cr}\sim 10^{-3.8} \, Z_\odot$ \citep{Schneider02, Schneider03, Schneider12} and is characterized by a "top-heavy" IMF with masses in the range [$10\msun-300\msun$] and a characteristic mass $m_{\rm ch}=20 \msun$. 
Conversely, Pop~II stars form out of chemically enriched gas ($Z>Z_{\rm cr}$) following a standard, Salpeter-like, IMF (approximated by a Larson IMF with characteristic mass $m_{\rm ch}=0.35\msun$) in the mass range [$0.1\msun-100\msun$].

In low-efficiency starburst, when the total stellar mass formed in $M_{\rm star}< 10^6 \, \rm M_\odot$, the intrinsic top-heavy Pop~III stellar IMF is stochastically sampled, randomly extracting single stars from the [$10\msun-300\msun$] mass range until the cumulative value of $M_{\rm star}$ is reached.

\subsection{Black hole seeds}\label{sec:seeds}

Following \cite{Valiante16}, BH seeds form under conditions set by the efficiency of metal and dust enrichment and by the intensity of the far UV radiation.

Depending on the random sampling of the IMF described above, light
seeds form in both mini-halos and atomic cooling halos by the collapse of $[40\msun -140 \msun]$ and [$260\msun-300\msun$] Pop~III stars (consistent with the existence of a pair instability mass gap). 
The resulting BHs (i.e. the collapsed remnants) are as massive as their progenitors, assuming non-rotating primordial stars, for which no mass loss is expected \citep{Heger02}.
Only the most massive BH of each population is assumed to settle in the galaxy center. 

In our seeding prescription, heavy BH seeds of $10^5\msun$ form in metal poor ($Z<Z_{\rm cr}$), atomic cooling halos, when the cumulative Lyman Werner (LW)  emission (from stars and accreting BHs in all galaxies), $J_{\rm LW}$, becomes larger than a critical threshold 
$J_{\rm cr}\equiv300\times 10^{-21} \, \rm erg\, s^{-1} Hz^{-1} cm^{-1}sr ^{-1}$ \citep[for a discussion see][and references therein]{Valiante17review}. 

The subsequent growth of nuclear BHs is driven by accretion of gas and mergers with other BHs. To describe the gas accretion rate we adopt the Bondi-Hoyle-Lyttleton formula, re-scaled by a factor $\alpha_{\rm BH}$ that accounts for the higher central densities around BHs, as required by sub-grid prescriptions adopted in SAM and in large-volume numerical simulations \citep[e.g.][]{DiMatteo05, BoothSchaye09}.
In addition, we assume that the computed BH accretion rate can not exceed the Eddington limit \citep[see][for details]{Valiante14}.

\subsection{Stellar and black hole feedback}
After each star formation episode, the galaxies ISM is polluted with metals and dust produced by supernovae (end products of main sequence stars of $10\msun$ to $40\msun$) and Asymptotic Giant Branch (AGB) stars (with initial mass of $1\msun $ to $8\msun$).The injection of fresh metals and dust produced by stars is regulated by the stellar lifetimes and depends on the initial mass and metallicity of the stars. We follow dust cycling in the two-phase ISM by accounting for SN shocks destruction in the hot, diffuse medium and grain growth in cold, dense molecular clouds \citep[see][for details]{Valiante14, deBennassuti14}. 
The stellar products can then be ejected out of the ISM, on scales larger than the halo virial radius. The energy released by star formation and BH accretion couples with the gas, heating and accelerating it.

We describe mechanical feedback by means of energy-driven winds: galaxy-scale gas outflows are launched from the galaxy polluting the IGM with metals and dust. In our models we assume that a fixed fraction of the energy deposited by SN explosions and BH accretion, $\epsilon_{\rm w,SN}=2\times 10^{-3}$ and $\epsilon_{\rm w,AGN}=2.5\times 10^{-3}$, respectively, drives the massive gas outflows \citep[see][for details]{Valiante16}.

We compute the time-dependent cumulative LW radiation, $J_{\rm LW}$ coming from all the emitting source, stars and active galactic nuclei \citep[AGN][]{Valiante16}. At each redshift, this can be considered as the background radiation permeating a comoving volume\footnote{This is the volume of the $10^{13} \, \rm M_\odot$ DM halo computed at the turn-around radius.} of $50$ Mpc$^3$\citep[see discussion in][]{Valiante17review, Valiante18b}.

\section{The dynamics of Binary Black holes}\label{sec:BBHs}\label{sec:BBHdynamics}

A description of BH dynamics in cosmological frameworks has been included, with different approaches, in several semi-analytical models so far \citep[e.g.][]{Volonteri03, Barausse12, Klein2016, Bonetti19, Katz2020}, and, recently, in few large-scales simulations 
\citep[associating time delays to BBHs in post-processing; see e.g.][]{2017MNRAS.464.3131K, Volonteri2020}.

In our previous models we assumed that during major mergers\footnote{In this paper major mergers refer to interacting DM halos with mass ratios greater than 1:4.} the BHs coalesce instantaneously as their hosts merge. In particular, in \citet[][]{Valiante16} BHs coalesce right away, over the merger tree time interval (that is typically of a few Myr), while in minor mergers the most massive BH remains in the center of the newly formed galaxy and the less massive is considered as a satellite and its evolution is no longer followed.

However, BH coalescences occur with a time delay compared to the typical time of the galaxy merger \citep{Colpi14}.
GW emission drives the inspiral on timescales of less than $\lesssim$ Gyr only when the two BHs reach relative separations of milliparsecs or smaller, depending on the binary mass, mass ratio and orbital eccentricity. 

During halo mergers, the two nuclear BHs can be driven to such minuscule galactic  distances 
by DM/stellar and gas dynamical torques that control their sinking from the kpc scale downward. 
Hence, the formation of BBHs in halo mergers and their hardening on timescales shorter than the cosmic time is an open and challenging multi-scale problem (see Section \ref{sec:discussion} for a discussion).

Within  \textsc{GQd}, we introduce a simplified treatment of BH dynamics, encompassing light and heavy seeds and massive BHs, by attributing to triple interactions the role of taxing BHs down to coalescence.\footnote{Actually, other physical mechanisms can influence the evolution of BHs, both before and after the pairing. See Section~\ref{sec:discussion} for a discussion.} This is motivated by the high incidence of multiple mergers among DM halos occurring at high redshift and traced by \textsc{GQd}. We adopt the model by \cite{Bonetti16,Bonetti18b,Bonetti18a} who carried out a large suite of numerical simulations with a three-body Post-Newtonian code describing the mutual interaction among BHs over a wide range of masses, mass ratios and orbit initial conditions, framed in spherical galactic potentials \citep{Bonetti16}. 
Multiple BH encounters provide a viable solution to the so-called final parsec problem (i.e. the stalling of binaries at separations below $\sim$ pc) when all other shrinking mechanisms are not efficient \citep{Bonetti18b} and are expected to have an important role in SMBH evolution (as in our model) as well as on future GW detections in particular in the LISA band \citep{Bonetti19}. 

Hereon,  we assume that a Keplerian BH binary forms promptly in a major halo merger, and that it is dragged in the nuclear region of the newly formed halo where it stalls until it interacts with a third incoming BH, called an intruder. 
This implicitly assumes that, within at most a few Myrs, the characteristic (redshift-dependent) time interval of our simulations, DM/stellar/gas dynamical friction is effective in forming a binary during the so called pairing phase, when the BHs sink as individual masses inside the halo merger remnant \citep{Begelman80}.  
Indeed a general expectation is that, at least when the two merging galaxies have mass ratios $>0.05-0.1$, dynamical friction efficiently drags the BHs from the outskirts towards the center of the newly formed galaxy within about a few million (up to a billion) years \citep{Mayer07,Callegari09,Van2014, Capelo2015,Khan16,Biava19}.

Subsequently, the intruder dragged by a third incoming DM halo can then interact with the binary via chaotic strong triple encounters or by Kozai-Lidov evolution following the formation of a bound hierarchical triplet \citep{kozai1962, Lidov1962,Bonetti16, Bonetti18a}.

In our model the fate of a triplet is defined on the basis of the statistical study presented by \citet{Bonetti18a}.  We use their results to distinguish triplets (potentially) leading to coalescence from systems that would never do (as their associated/computed merger timescale is longer than the Hubble time at $z=0$). In this way we are accounting for the (global) efficiency of triple interactions in driving BH mergers, limiting the fraction of events. 
Quadruple encounters are reduced to a three-body problem by means of the ejection of the lightest BH and iterated as triple systems \citep[see][for details]{Bonetti18b}.

In a triple encounter, the pairs that eventually coalesce are selected on the basis of the merger fractions and relative occurrence probabilities computed by \citet[][]{Bonetti18a}.
We assign a probability to any pair of BHs in a triple merger and randomly extract the outcome of the interaction by interpolating through their model grid of primary BH masses, $m_1$, inner and outer mass ratios, $q_{\rm in}$ and $q_{\rm out}$ 
\footnote{The probabilities of the closest grid point are assigned to triplets whose parameters are outside the range sampled by \citet[][]{Bonetti18a}.}.
These same properties also define the merger timescale of each system. However, in our SAM we adopt a simplified assumption: in successful triplet-induced merger the two BHs coalesce within the characteristic simulation time interval (up to few Myr). This assumption implies that BBH merger times are determined mainly by the sequence (rate) of BH-seeded halo-halo encounters within a merger tree, rather than by dynamical processes. We will discuss this point in Section~\ref{sec:discussion}.

To summarize, in our model BH mergers are triggered only via triplets formation and triple interactions and have two possible outcomes: {\it (i)} the "instantaneous" coalescence of any two BHs; {\it (ii)} the formation of a so-called "left-over" binary (no merger), with the ejection at larger scales of one of the involved BHs (usually the lighter). 

\begin{table*}
	\begin{center}
	\caption{Properties of selected quasars and main free parameters of \textsc{GQd} models.}
	\label{tab:parameters}
	\begin{tabular}{ccccccc} 
		\hline
		Object & $z_{\rm QSO}$ & SFR ($\rm M_\odot/yr$) & $\log (M_{\rm SMBH}/\msun)$ & $\epsilon_{\rm quiesc}$ & $\alpha_{\rm BH}$ & $\epsilon_{\rm AGN,w}$\\
		\hline
		J1148 & 6.4 & 100-1000 & $9.5^{+0.3}_{-0.2}$ & 0.1 & 110 & $2\times 10^{-3}$\\
		J2345 & 2.0 & 50-330 & $9.47\pm{0.3}$  & 0.5 & 50 & $2\times 10^{-3}$\\
		PDS456   & 0.2 & 30-80 & $9.4\pm{0.17}$  & 0.5 & 50 & $2\times 10^{-3}$\\
		\hline
	\end{tabular}
	\end{center}
\end{table*}
\begin{figure*}
	\includegraphics[width=\columnwidth]{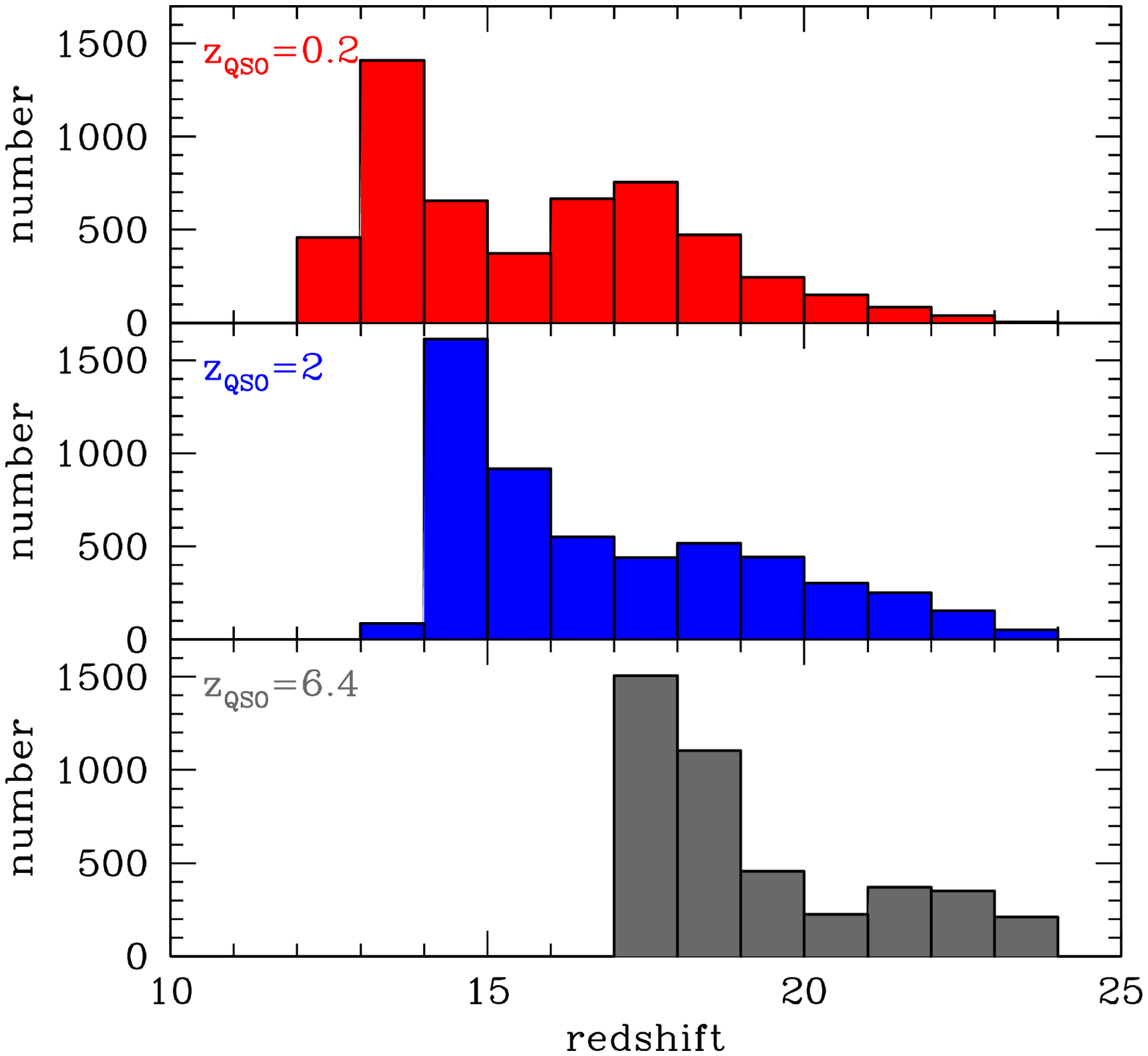}
	\includegraphics[width=\columnwidth]{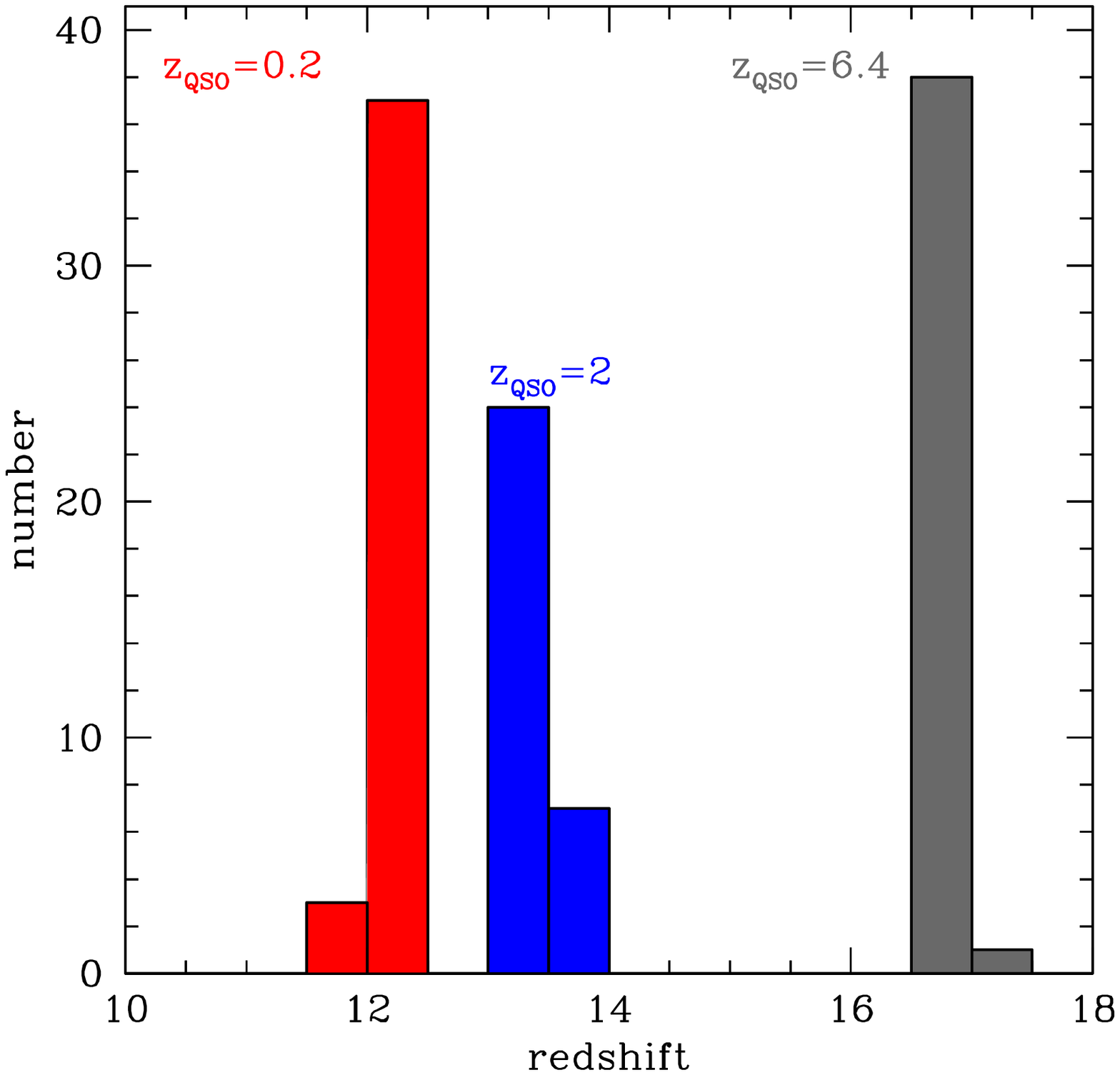}
    \caption{Redshift distribution of the number of light (left panel) and heavy (right panel) seed BHs forming along the "fiducial" merger history of quasar J1148 outshining at $z_{\rm QSO}=6.4$ (grey histograms), J2345 at $z_{\rm QSO}=2$ (blue histograms) and PDS456 at $z_{\rm QSO}=0.2$ (red histograms).}
    \label{fig:seeds}
\end{figure*}

\section{The emergence of binary Black holes}\label{sec:QSOs}
As mentioned in $\S$\ref{qso-evolution}, the model reconstructs the formation histories of three luminous quasars at $z_{\rm QSO}=6.4$ $2.0$ and $0.2$. We choose as proto-typical objects for the three redshifts the quasars J1148 at 
$z_{\rm QSO}=6.4$ \citep{Fan01}, SDSS J2345+1104 at $z=2$ \citep[hereafter J2345 ]{Shen11, Schulze19}, and PDS 456 at $z=0.2$ \citep[PDS456 hereafter]{Nardini15, Bischetti19}.

We model the evolution of J1148, J2345 and PDS456 performing, for each of them, ten independent simulations adopting the set of model parameters described in Table~\ref{tab:parameters}. These are tuned to reproduced the observed SMBH
mass and host galaxy physical properties \citep[see][for more details]{Valiante11,Valiante14,Valiante16}.

To investigate the emergence of BBH populations across the cosmic epochs (in our cosmological framework), we select one "fiducial"  simulation (out of the 10 performed) for each template quasar. In particular, in what follows we show the results of the simulation that provides a global SMBH evolution that best matches the corresponding simulation-averaged predictions.

\subsection{From seeds to binaries along a merger tree} \label{sec:seedsToBBHs}
In Fig.~\ref{fig:seeds} we show the distributions of light seeds (on the left panel) and heavy seeds (on the right) as a function of their formation redshift. Grey, blue and red histograms refer to  quasars J1148, J2345 and PDS456, respectively. 

The number of seeds as well as the shapes of the histograms are similar for the three quasars. 
The bi-modal distribution of light seeds reflects the properties of Pop~III star forming halos. At early times they are mainly mini-halos where star formation is dramatically limited by radiative feedback (H$_2$ photo-dissociating radiation in particular). At later epochs, Pop~III stars (and thus light seeds) instead mainly form in atomic cooling halos which are less affected by the presence of external UV radiation \citep[see][for a more detailed description]{Valiante16}. 

On the other hand, heavy seeds form only in atomic cooling halos and the environmental conditions required by the direct collapse BH formation scenario (sub-critical metallicity and a super-critical illuminating LW radiation, see Section~\ref{sec:seeds}) are met only over a very limited period of time and by a limited, very low, number of halos within our merger trees.

A total of 39 (31 and 40) heavy and 4228 (5327 and 5319) light seeds are formed along the assembly history of J1148 (J2345 and PDS456, respectively). 
In all cases, light seeds form in larger numbers at very high redshift ($12<z<30$) and over a longer period of cosmic evolution than heavy seeds, which are rarer (with relative fraction $\sim 1\%$) and form for a shorter period of time at slightly lower redshift ($z\sim 12 - 17$, depending on the considered system).

In-situ and/or external pollution determines the end of the seed (and Pop~III stars) formation era: as soon as all the galaxies have been enriched above the critical metallicity threshold ($Z_{\rm cr}=10^{-3.8} \rm Z_\odot$), the transition to the Pop~II star formation regime is completed. This critical level is reached, on average, at  $z\sim 16$, $13$ and $12$ for quasar J1148, J2345 and PDS456, respectively. Below this redshift, light and heavy seeds no longer form.

\begin{figure}
    \includegraphics[width=\columnwidth]{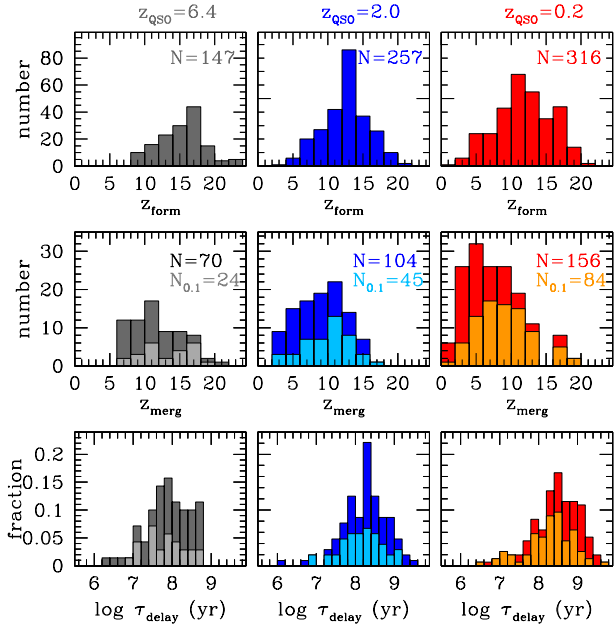}
    \caption{\textit{Upper panels:} number of BBHs as a function of their formation redshift, $z_{\rm form}$, in quasar models for J1148 (gray histogram on the left), J2345 (blue histogram, in the middle) and PDS456 (red histogram, on the right). Labels in each panel indicate the total number of binaries that form, summed over all redshifts. \textit{Central panels:} distribution of the number of triplet-driven merging BHs at their merger redshift, $z_{\rm merg}$, for the same three quasars. The total number of BH coalescences are labelled in dark in each panel. \textit{Bottom panels:} distribution of BBH merger time delays, $\tau_{\rm delay}$ (or lifetimes,  see text for details). In central and bottom panels lighter colours show distributions for merging binaries with mass ratio $q\geq 0.1$.  }
    \label{fig:binariesHist}
\end{figure}

For each of the three simulations, Fig.~\ref{fig:binariesHist} shows the number of BBHs at their formation redshift, $z_{\rm form}$ (upper panels), and the number of merging BBHs at their coalescence redshift $z_{\rm merg}$ (central panels). The latter is the redshift at which a triplet BH system forms, leading to the prompt coalescence of a BBH, according to the physical prescriptions described in Section \ref{sec:BBHs}.

In the upper panel of Fig.~\ref{fig:binariesHist} we show that  147 (257, 316) binaries form over the simulated cosmic time, $\sim 900$ Myr ($3$ and $11$ Gyr) for quasar J1148 (J2345 and PDS456, respectively).\footnote{Each simulation of a quasar is characterized by its peculiar number of seeds and halo major mergers. For example, the number of heavy seeds that form across cosmic times, along a merger tree, can vary from a few up to few tens, depending on the specific simulation, mirroring the relative efficiency of chemical and radiative feedback in each history \citep[see][for a discussion]{Valiante16}. However, we find that the redshift intervals over which the seeds and BBHs form and merge, as well as the merger timescales distribution, are very similar, i.e. do not vary much, among the different simulations of a given quasar.}

The histograms in the bottom panels of Fig.~\ref{fig:binariesHist} show the distribution of the  {\it delay times}  to coalescence, $\tau_{\rm delay}$. In our model, the merger timescale of two BHs, following halo assembly, corresponds to the time elapsed from the formation of the $i-$th binary down to coalescence, driven by a successful multiple BH interaction, involving that binary, i.e.  $\tau_{\rm delay,i}=t(z_{\rm merg,i}) - t(z_{\rm form,i})$. 
The mean values of the delay time distributions are of the order of $\sim 150$, $360$ and $590$ Myr, respectively in the simulations of quasars like J1148 ($z_{\rm QSO}=6.4$), J2345 ($z_{\rm QSO}=2$) and PDS456 ($z_{\rm QSO}=0.2$).

These delays correspond to the typical timescales of  triple halo interactions, each hosting a nuclear BH. 
Two additional delay times should be considered: the formation timescale of the binary system (i.e. the time required for the nuclear black holes of the merging halos to reach the center of the newly formed system and to dynamically pair) and the time required for the system to coalesce. These timescales are not considered in the present study. A detailed discussion will be presented in Section~\ref{sec:discussion}.

Finally, histograms drawn in lighter colours in the central and bottom panels of Fig.~\ref{fig:binariesHist} represent the distributions of BBHs with a mass ratio $q \ge 0.1$, whose gravitational wave emission will be analyzed in the next Section. 

\subsection{Merging black holes in the ET and LISA frequency domains}
\begin{figure}
   \centering
	\includegraphics[width=8.7cm]{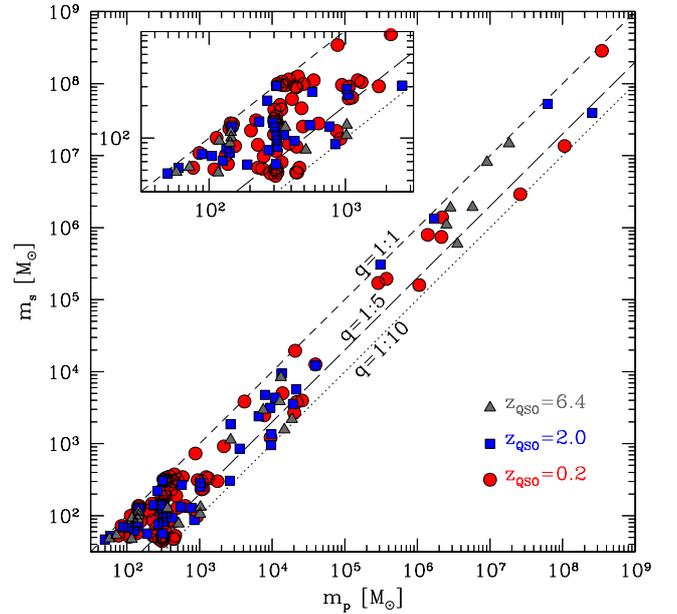}
    \caption{Mass of the primary ($m_{\rm p}$) and secondary ($m_{\rm s} < m_{\rm p}$) components of merging BBHs with mass ratio $q=m_{\rm s}/m_{\rm p}\geq 0.1$, 
    formed along the evolutionary history of the three quasars: J1148 at $z_{\rm QSO}=6.4$ (grey triangles), J2345 at $z_{\rm QSO} =2$ (blue squares) and PDS456 at $z_{\rm QSO}=0.2$ (red circles). Dashed, log-dashed and dotted lines mark secondary over primary mass ratios equal to 1:1, 1:5 and 1:10, respectively. The inserted box on the top left zooms on $<10^3 \, \rm M_\odot$ binaries.}
    \label{fig:binariesMass}
\end{figure}

In this section we describe the properties of coalescing BBHs extracted from \textsc{GQd} for each of the three quasars. Then, we discuss their detectability in the ET high frequency and LISA low frequency domains.

The primary (most massive) and secondary BH masses in merging BBH systems are shown in Fig.~\ref{fig:binariesMass}. Given the wide mass interval probed by \textsc{GQd}, halos are found to host dual/multiple black holes with mass ratios as small $10^{-2} - 10^{-4}$ for which we could not follow their as yet unknown (likely erratic) dynamics. 
Binaries with such small mass ratios might never form as a consequence of the long dynamical friction timescale \citep[e.g.][]{Dosopoulou17}. For this reason, we do not include these systems in our analysis and the figure reports BBHs with mass ratio $q\equiv m_{\rm s}/m_{\rm p}>0.1$, which
cover almost uniformly the $0.1 \leq q \leq 1$ interval.

\begin{figure*}
    \centering
    \includegraphics[width=0.8\textwidth]{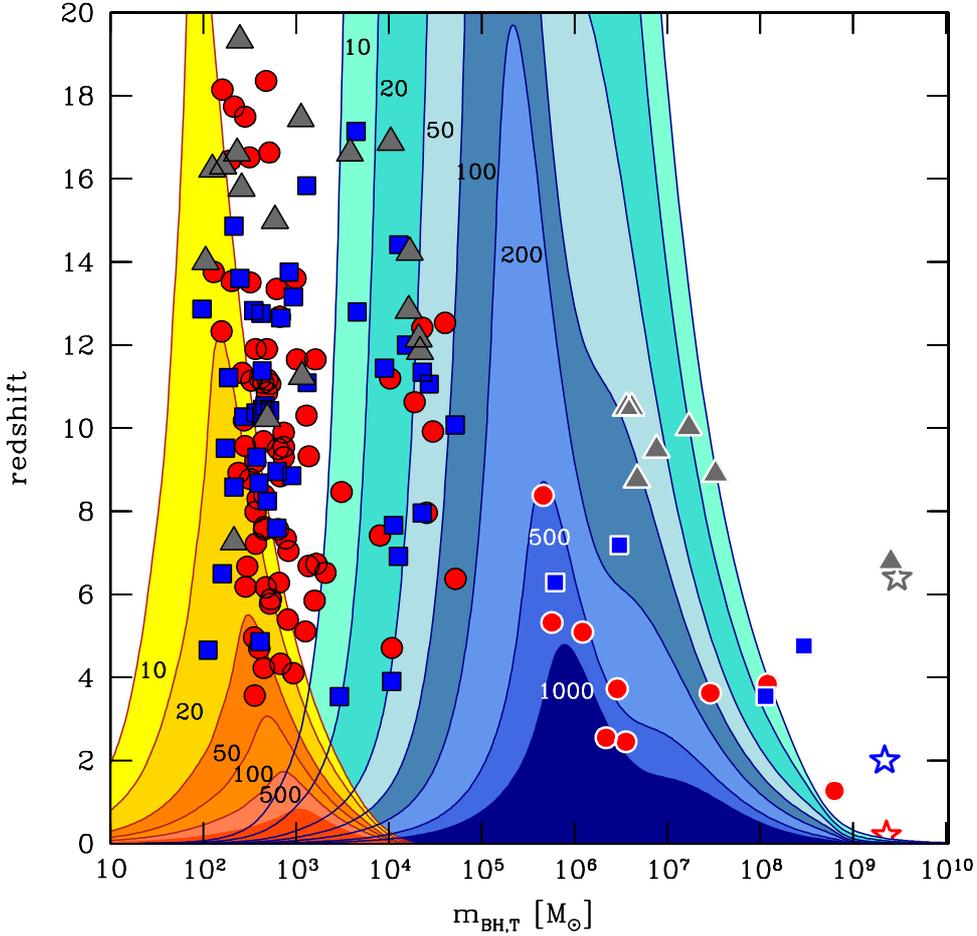}
    \caption{Distribution of BBH coalescence events in the redshift $z$--$m_{\rm BH,T}$ diagram. Data points describe cosmologically-driven BH mergers with mass ratio $q\geq 0.1$, triggered only by triple interactions among galaxy halos. Grey triangles, blue squares and red circles denotes tha total mass and redshift of the coalescences extracted from the simulation of a $10^{13}\msun$ over-density, forming a $\sim 10^{9}\msun$ SMBH at $z_{\rm QSO}= 6.4$, $2$ and $0.2$ (represented with stars in the plot). Symbols with white edges indicate mergers involving at least one heavy seed. Color-coded areas represent lines of constant S/N ratios for ET (yellow/red) and LISA (azure/blue) computed for non spinning binaries assuming a mass ratio $q=0.5$, which corresponds to the mean value of the merging binaries extracted from our samples. The ensemble of the color-coded areas for a given detector is often referred to "waterfall" plot and provides averaged values of the S/N ratio at which a GW source is detected. 
    }
    \label{fig:WFplots}
\end{figure*}

In Fig.~\ref{fig:WFplots} we show the distribution of BH mergers in the $z$--$m_{\rm BH,T}$ plane, where $m_{\rm BH,T}$ is the total mass of the binary in the source rest frame. 
Different symbols/colors pinpoint cosmologically-driven BBH coalescences triggered by triple interactions that occur during the assembly of the three 
simulated quasars: J1148 (grey triangle-$z_{\rm QSO}=6.4$), J2345 (blue square-$z_{\rm QSO}=2$) and PDS456 (red circle-$z_{\rm QSO}=0.2$).
Data points with white edges indicate mergers involving at least one heavy seed.

\begin{figure*}
    \centering
    \hspace{-0.5cm}
    \includegraphics[width=7cm]{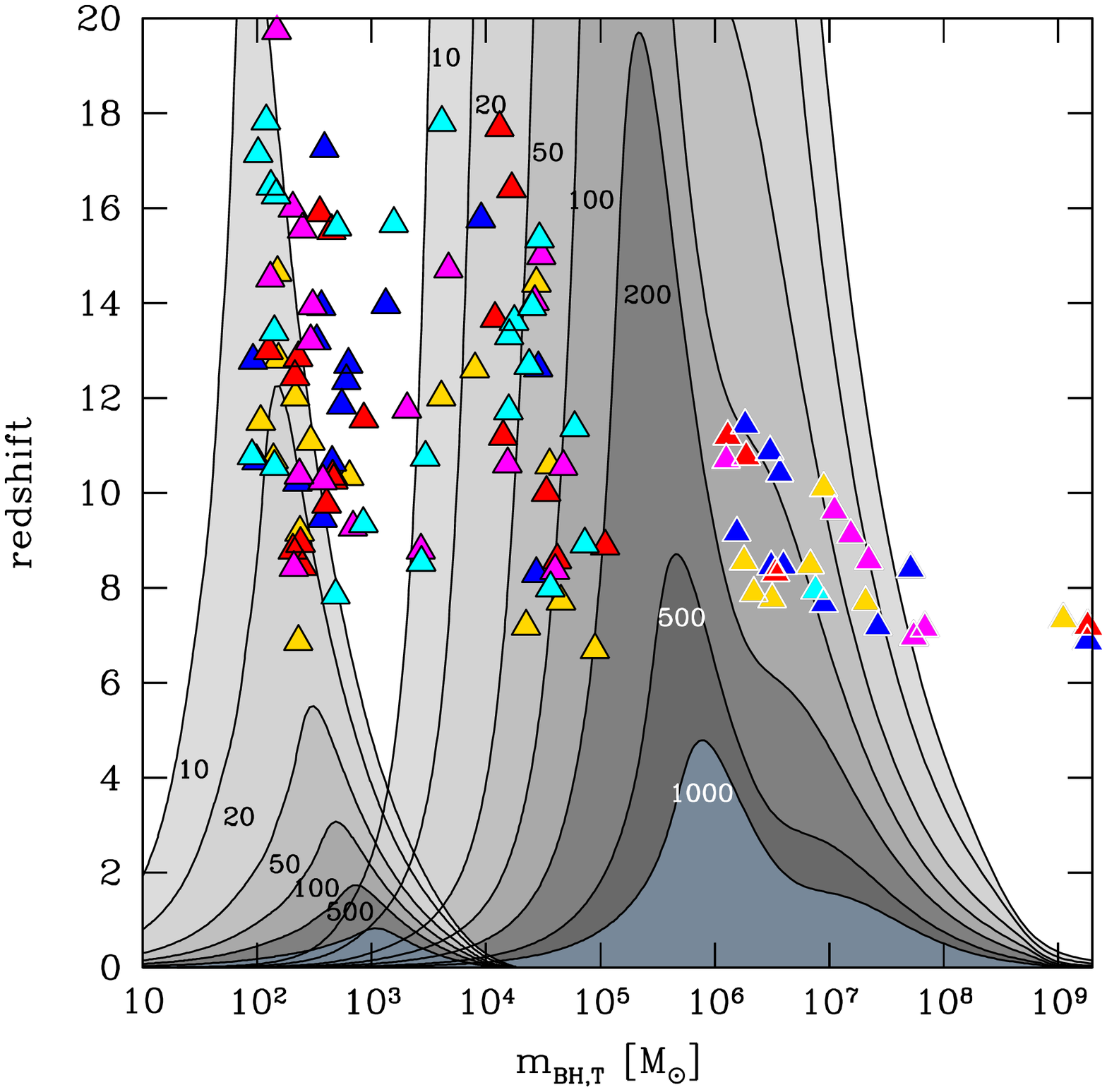}
    \hspace{-1.6cm}
    \includegraphics[width=7cm]{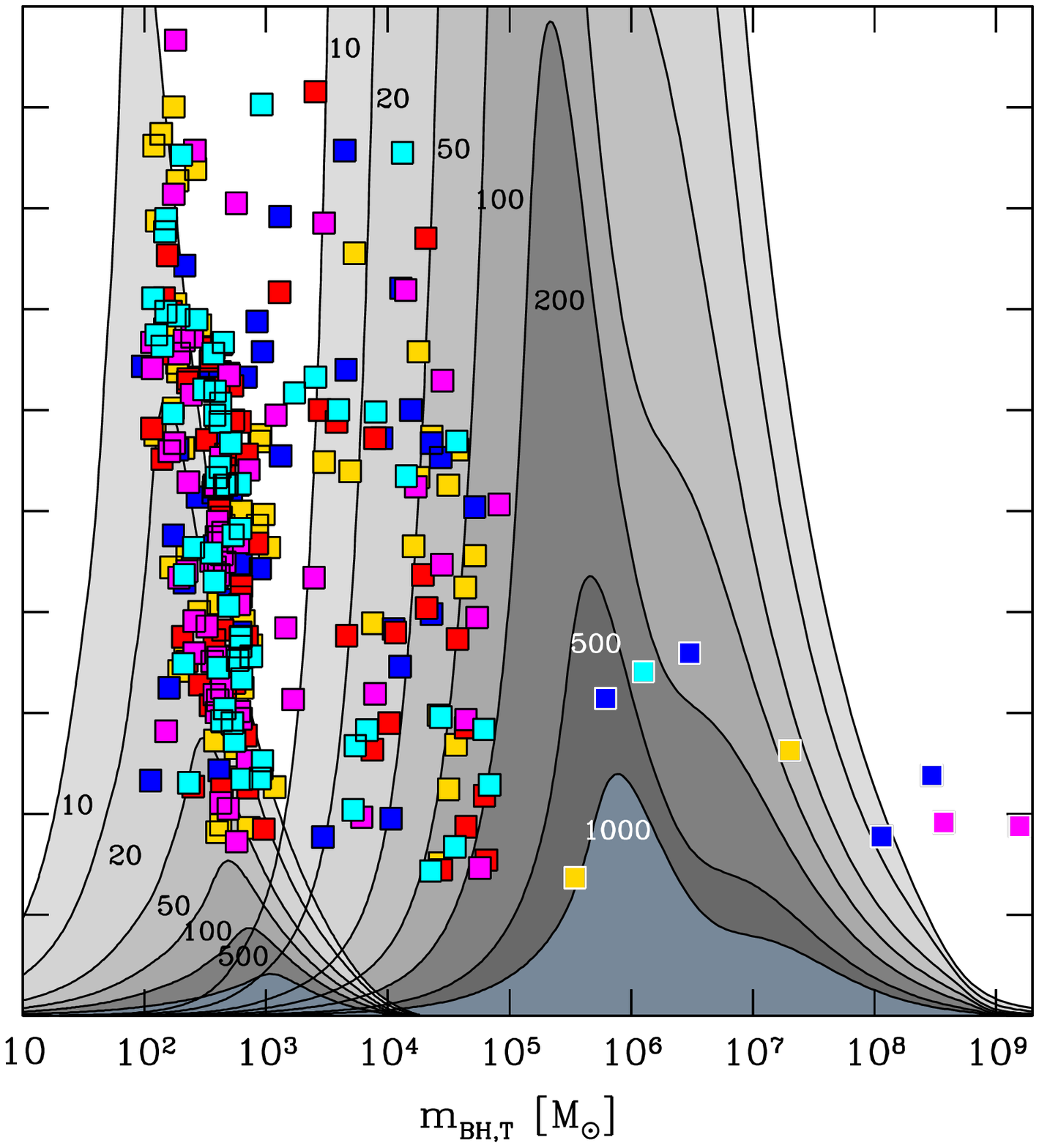}
    \hspace{-1.6cm}
    \includegraphics[width=7cm]{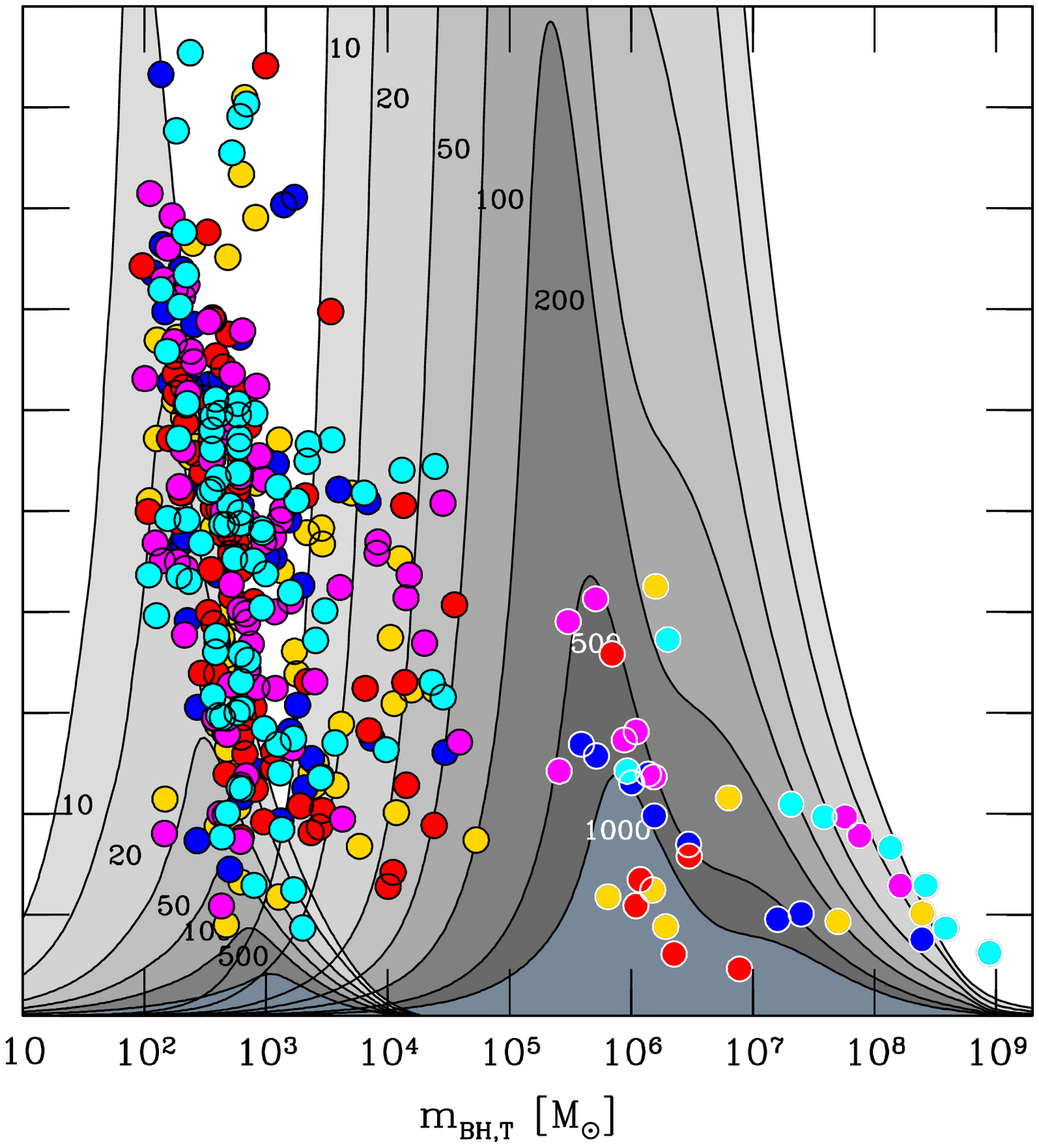}
    \caption{Same as in Fig.~\ref{fig:WFplots} but comparing 5 simulations of each quasar at $z_{\rm QSO}=6.4$ (left panel), 2 (central panel) and 0.2 (right panel). In all panels different colors indicate BH coalescences extracted from a given merger tree simulation of the same considered quasar. The ET and LISA Waterfall plots are shown in different shades of gray, for simplicity.}
    \label{fig:dispersion}
\end{figure*}

Overlaid in Fig.~\ref{fig:WFplots} are contour lines of constant Signal-to-Noise (S/N) ratio computed using the ET-D sensitivity curve by \cite{Hild2011} for ET, 
and that of \cite{Robson_2019} for LISA.  The IMRPhenomC \citep{Santamaria10} gravitational waveform family is used to compute the strength of the signal assuming non spinning BHs, which includes only the 22 quadrupolar mode. The ensemble of color-coded areas for a given detector is often referred to as "waterfall" plot that provides values of the S/N ratio at which a GW source would be detected, averaged over the source's sky position, the binary-inclination and GW-polarization angles.\footnote{A S/N threshold between 5 and 10 is customarily taken as detection threshold for any GW event. Here we consider $\rm S/N=10$ as the detection threshold.}
\footnote{ CE and ET will be part of a network of detectors that will enlarge the GW cosmic horizon.  Although they will have comparable sensitivities, ET will be more sensitive below 10Hz, with CE more sensitive at higher frequencies.  Consequently, ET will have better sensitivity to higher mass mergers ($\gtrsim 100\msun$), with CE being more sensitive at lower masses \citep[$\lesssim 20\msun$; ][]{HallEvans2019}.  As the focus of this study is on binaries above $100\msun$, we only show the sensitivity of ET in  Figures \ref{fig:WFplots}, \ref{fig:dispersion} and \ref{fig:GWsummary}.}
 
 The figure shows that both observatories shall have the capability of detecting GWs from coalescences occurring at 
redshifts as large as $z\sim 15$ (and even beyond for a narrower interval of masses) letting us
explore the epochs of seed formation and growth.  But not only that. Coalescence events are found to spread over a much wider range in redshift and mass: down to $z\simeq 2$ and up to a few $ 10^7\msun$.  The lack of mergers at very low redshifts is a consequence of our model assumptions. 

In more detail,  Figure \ref{fig:WFplots} shows how densely populated are the two GW windows during the cosmic assembly of our simulated quasars. The fastest evolution is associated to the $z_{\rm QSO}=6.4$ quasar. Here the galaxy halos  and black holes evolve at a rapid pace and the associated GW events drift away from the ET bandwidth swiftly, most of them transiting across the deci-Hz window \citep{Sato2017,Voyage2019} already at $z>12$. 
For this quasar model (J1148), a few coalescences of BBHs involving pairs of light seeds would be visible in the ET band at $z = 14  - 16$. However, most of the events involving BHs grown from light seeds occurring mainly at $z \geq 12$ would be visible in the LISA band when the BHs have achieved masses of $\sim 10^4 \msun$, due to efficient gas accretion in the environments.
We have to wait until redshift $z \sim 10-11$ 
to see a $q>0.1$ merger involving at least one BH grown from a heavy seed. 
The bulk of these heavy-seed mergers (3 involving two BHs grown from heavy seeds and 4 with BHs pairs grown from a light and a heavy seed) occurs between $8<z<11$, when the Universe is only 600 Myrs old.

Despite our results are based on a limited number of trials, we do find that, generically, mergers (with $q>0.1$) involving BHs which originate from heavy seeds  appear in the LISA band when the original seeds have already increased their mass by gas accretion up to $M_{\rm BH} \geq 10^6 \msun$.
A similar trend is observed also in the BBHs formed along the evolution of the $z_{\rm QSO}=2$ and $0.2$ quasars. 
In the latter case, $\sim 20\%$ of all detectable mergers involving heavy seeds are found in the mass range $10^5<m_{\rm BH,T}/\msun <10^6$. 
We warn, however, that if heavy seeds were to form with a wider mass spectrum than considered in our model, extending from less than $10^4\msun$ up to a few $10^5\msun$, the mid region of the LISA band would also be populated of events. 

In our model, the assembly of a quasar at redshift $z_{\rm QSO}=6.4$ constrains the flow of data points across the $z-m_{\rm BH,T}$ plane. It acts as a terminal point of the cosmological evolution of BH seeds.
Detecting a coalescence at redshift as large as $z\sim 10-14$ with ET and a coalescence just on the edge of the left side of the LISA waterfall plot at adjacent redshifts would provide the first evidence that light seeds form and grow via accretion in high-$z$ gas-rich environments, and are dynamically paired in coalescing binaries during galaxy mergers.

Yet, the detection of these events is challenging. In ET, they mainly lie in the declining (right) side of of the waterfall envelope and are characterized by low S/N ratios \citep{Kalogera19}.  Here the portion of the detected GW signals traces only at most 1-2 cycles of the inspiral, and the merger and ringdown. On the other hand, the coalescence of light seeds grown up to masses of $10^3-10^4\msun$ at $z$ larger than $10$ lie in the rising side of the waterfall envelope in the LISA band, and are far louder GW sources, with S/N ratios in the range 10--50. The GW signal takes now the shape of a nearly adiabatic inspiral, as their merger falls in the deci-Hz window.  

Coalescences that involve heavy seeds in the LISA band at $z\sim 10$ are at the edge of the declining side of the waterfall plot, and their GW signal (with S/N ratios in the $[10-100]$ interval) is dominated again  by few cycles in the  inspiral, and by the  merger and ringdown phases. 
 
Due to the incompleteness of our modelling, we cannot exclude presence of evolved seeds of $10^5\msun$ up to a few $10^6\msun$ which will be observable as high S/N GW sources in LISA.

The assembly histories of the two remaining quasars have as anchor points of the simulations two lower redshift systems, and as a consequence coalescences are distributed over a wider redshift interval, implying the appearance of lower-$z$, louder GW sources, both in the ET and LISA frequency domains.  

Considerations similar to those discussed for the highest redshift quasar simulation apply here. But here we clearly see that, besides the population of BBHs swiftly transiting to higher masses (to enable the formation of a SMBH of $\sim 10^9 \msun$), there exist a lower redshift population, that we call "starved" binary seeds, with masses in the range between $100 \msun$ and a few $10^3 \msun$. 
These systems are hosted by halos where seeds were unable to grow or that grew only marginally, filling the middle weight mass range. Also, the number of BH mergers increases, reflecting the larger number of progenitor halos (and thus halo-halo coalescences) in the merger trees of the lower$-z$ simulated quasars (see Section~\ref{sec:DMhalo}): while we witnessed 24 mergers in the $z_{\rm QSO}=6.4$ halo, in  the $z_{\rm QSO}=2$ and  $z_{\rm QSO}=0.2$ halos, we have 45 and 84 mergers, respectively with only a handful (less than $10\%$, on average) comprising heavy seeds.

\subsubsection{The impact of cosmic variance}
Although we analyze here a single simulation of each quasar, the findings discussed above do not dramatically depend on the selected simulation. 
The dispersion in the $z - m_{\rm BH,T}$ plane due to the choice of a specific merger tree simulation can be appreciated in Figure~\ref{fig:dispersion} where we collect BH mergers, with $q>0.1$, extracted from 5 realizations of each quasar. 

For the $z_{\rm QSO}=6.4$ quasar the choice of the merger history mainly affects the fraction of mergers (involving pairs of light seeds) that could be detected in the ET band at $z<12$. This varies from $18\%$ (magenta triangles) up to $\sim 35\%$ (cyan triangles); $24\%$ is found for the realization shown in Figure~\ref{fig:WFplots}).  
The redshift distribution of starved binaries extends towards lower redshifts (down to $z\sim 2$) when different formation histories are considered for the $z_{\rm QSO}=0.2$ quasar.
Within our model we can investigate the relative occurrence of binary coalescences involving BHs of different origins, along the quasar evolution history. The vast majority of BH mergers ($\sim 90\%$) involve pairs of light seeds, as they are more common than heavy seeds as shown in Figure~\ref{fig:seeds}. The fraction of mergers involving the two seed flavours is reported in Table~\ref{tab:fraction}.

{ 
The merging binaries in our models could  be "multi-band sources", i.e. sources that transit from the LISA low frequency domain (during their long-lived inspiral phase) to the ET/CE high frequency domain (merger and ringdown) if their lifetime in the LISA band is shorter than the nominal lifetime of the mission (4-10 years).  Joint  multi-band observations of the same event will be possible for $(10^2-10^4)\msun $ BBHs out to redshift $\sim 4-5$ \citep[e.g.][]{Jani2019}. Multi-band detections of distant lower (higher) mass binaries would be instead limited by the sensitivity for LISA  at frequencies around and above 0.1 Hz  (for ET/CE at frequencies around  and below 3 Hz).
We note here further that very few mergers in the $z_{\rm QSO}=0.2$ quasar model (right panel of Figure~\ref{fig:dispersion}) are predicted to be observable both in ET and LISA (none in the "fiducial" model shown in Figure~\ref{fig:WFplots}).
}
\begin{table}
    \centering
    \begin{tabular}{c|c|c|c|c}
       \hline
        & & single simulation & & \\
        \hline
        {\bf $z_{\rm QSO}$} &  {\bf $n_{0.1}$} & {\bf $f_{L-L}$} & {\bf $f_{L-H}$} & {\bf $f_{H-H}$}\\
        \hline
         6.4 & 24 & $71\%$ & $17\%$ & $13\%$ \\
         2.0 & 45 & $91\%$ & $7\%$  & $2\%$ \\
         0.2 & 84 & $89\%$ & $11\%$ & $0$ \\
         \hline
         \hline
         & & simulations-averaged & & \\
         \hline
         {\bf $z_{\rm QSO}$} &  {\bf $n_{0.1}$} & {\bf $f_{L-L}$} & {\bf $f_{L-H}$} & {\bf $f_{H-H}$}\\
         \hline
         6.4 & 27 & $86\%$ & $12\%$  & $2\%$ \\
         2.0 & 62 & $97\%$ & $2.4\%$ & $0.6\%$ \\
         0.2 & 78 & $89\%$ & $8.3\%$ & $2.5\%$ \\
         \hline
    \end{tabular}
    \caption{Statistical analysis of BH mergers: the number of BH mergers with mass ratio $q\geq 0.1$ ($n_{0.1}$) and the fraction of these coalescences involving pairs of light ($f_{L-L}$), light+heavy ($f_{L-H}$) and heavy ($f_{H-H}$) seeds. The upper table refers to the single "fiducial" realizations of each simulated quasar while mean values, averaged over 10 merger histories for each system, are reported in the bottom table.}
    \label{tab:fraction}
\end{table}

We further remark that coalescing stellar BBHs, relic of massive population III stars, could also form in situ \citep{Hirano2018, Sugimura2020}. These non cosmologically-driven mergers are not included in the figure, nor the population of binaries forming via dynamical captures in dense environment such as young star clusters \citep[e.g.][]{DiCarlo2020} and globular clusters \citep[e.g.][]{Rodriguez2016, Askar2017}
or in galactic fields via ordinary channels \citep[e.g.][]{Dominick2012, Dominick2013, Dominick2015, Mapelli2017, Mapelli2018, Mapelli2019}, particularly the most massive ones \citep{Schneider2017, Marassi2019, Graziani2020}.
We expect that these stellar BHs will preferentially fill the left corner of the ET waterfall plot, as shown later in Figure~\ref{fig:GWsummary}, extending out to the redshifts at which star formation started \citep{Santoliquido2020}.

\subsubsection{Event rates in the LISA and ET sky}
{ 
In principle, to compute the total number of detectable sources per year, namely the event rates in the LISA and ET band, we would need to simulate a large number of merger trees, spanning a wide range of parent halo masses and formation redshift (weighting each mass according to the expected halo mass function). 
Nevertheless, at $z\sim 0$ DM halos of $10^{13} \, \rm M_\odot$ are expected to be common, thus, using our $z_{\rm QSO}=0.2$ model as representative of an "average Universe" would provide a reasonable estimate of the merger rates\footnote{The $z_{\rm QSO}=6.4$ and $z_{\rm QSO}=2$ predictions would provide extremely incomplete estimates of the detectable event rates as, at those redshifts, DM halos of $10^{13} \rm M_\odot$ are instead the highest $\sigma$ mass density fluctuations, representative of highly biased regions of the Universe.}. 

To this aim, we first compute the intrinsic rates and chirp masses of the BH-BH mergers extracted from our 10 realizations of the $z_{\rm QSO}=0.2$ quasar model\footnote{Using the average comoving volume occupied by a typical $10^{13} \, \rm M_\odot$ halo ($\sim300 \, \rm Mpc^3$) to weight our intrinsic merger rates, we find a simulations-averaged total value of $\sim 83$ mergers per year ($\sim 93/$yr using the "fiducial" simulation presented in Fig.~\ref{fig:WFplots}).}. 

We use these information to generate a Monte Carlo sample of all mergers occurring in 100 years. Then the S/N of simulated binaries is determined using the IMRPhenomC \citep{Santamaria10} waveforms with the corresponding sensitivity curves adopted in Figure~\ref{fig:WFplots}. 
For each binary, we randomized over sky position, inclination and polarization in order to compute the fraction of detected sources. Assuming that only sources with ${\rm S/N} > 12 \, (8)$ can actually be detected by ET (LISA), we obtain a total of $11.25 \, (18.7)$ events per year.
The event rates in the LISA band are comparable to those obtained in other studies \citep[e.g.][]{Ricarte2018, Bonetti19, Dayal2019}.

{
Note however, that populating the Universe only with $10^{13} \, \rm M_\odot$ halos may lead to an overestimation of the number of events per year, when compared with merger rates weighted appropriately on the Press-Schechter (PS) halo mass function.
By extracting the merger rates for $10^{13} M_\odot$ halos and for the PS-weighted halo population from the model of \citet{Barausse2012},
we find the results to differ by a factor of $\lesssim 2.5$.}
Therefore, although crude, our estimate should be reliable within a factor of $\approx 2 - 3$.

Merger rates $\lesssim 1/$yr are instead obtained from the $z_{\rm QSO}=2$ and $6.4$ models, normalizing the intrinsic rates to the observed number density of bright quasars at those redshifts ($\sim 10^{-7}$ and $\sim 10^{-9}$ Mpc$^{-3}$, respectively). This suggests that $z=2$ and $z=6.4$ quasars would contribute only a small fraction to the overall observed rate.

We stress here that computing actual/realistic merger rates is not one of the goals of this work, but will be the focus of future, improved, studies.
}

\section{Observing the earliest accreting BHs with EM waves}\label{sec:EM}

\begin{figure*}
    \includegraphics[width=\columnwidth]{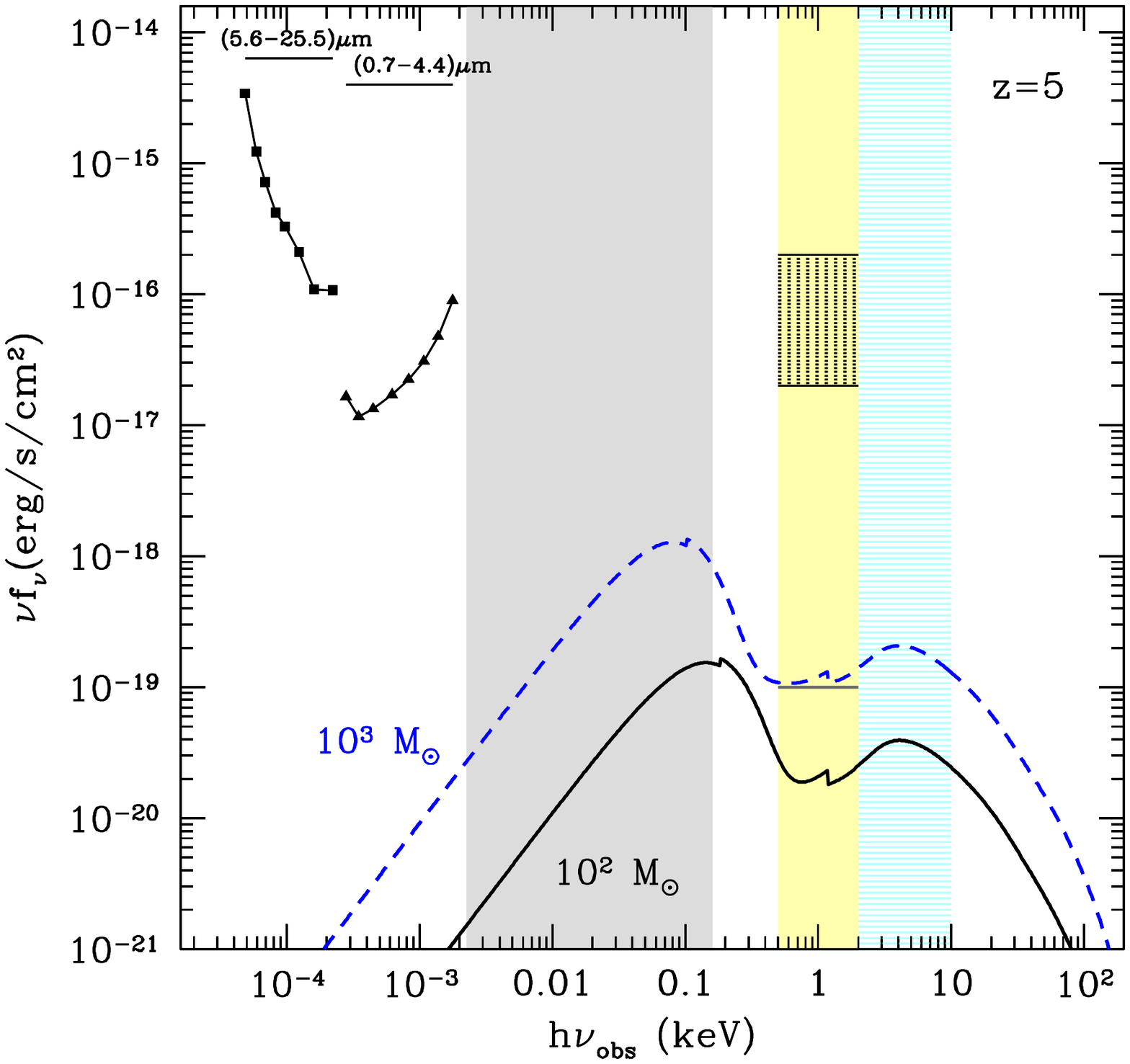}
    \includegraphics[width=\columnwidth]{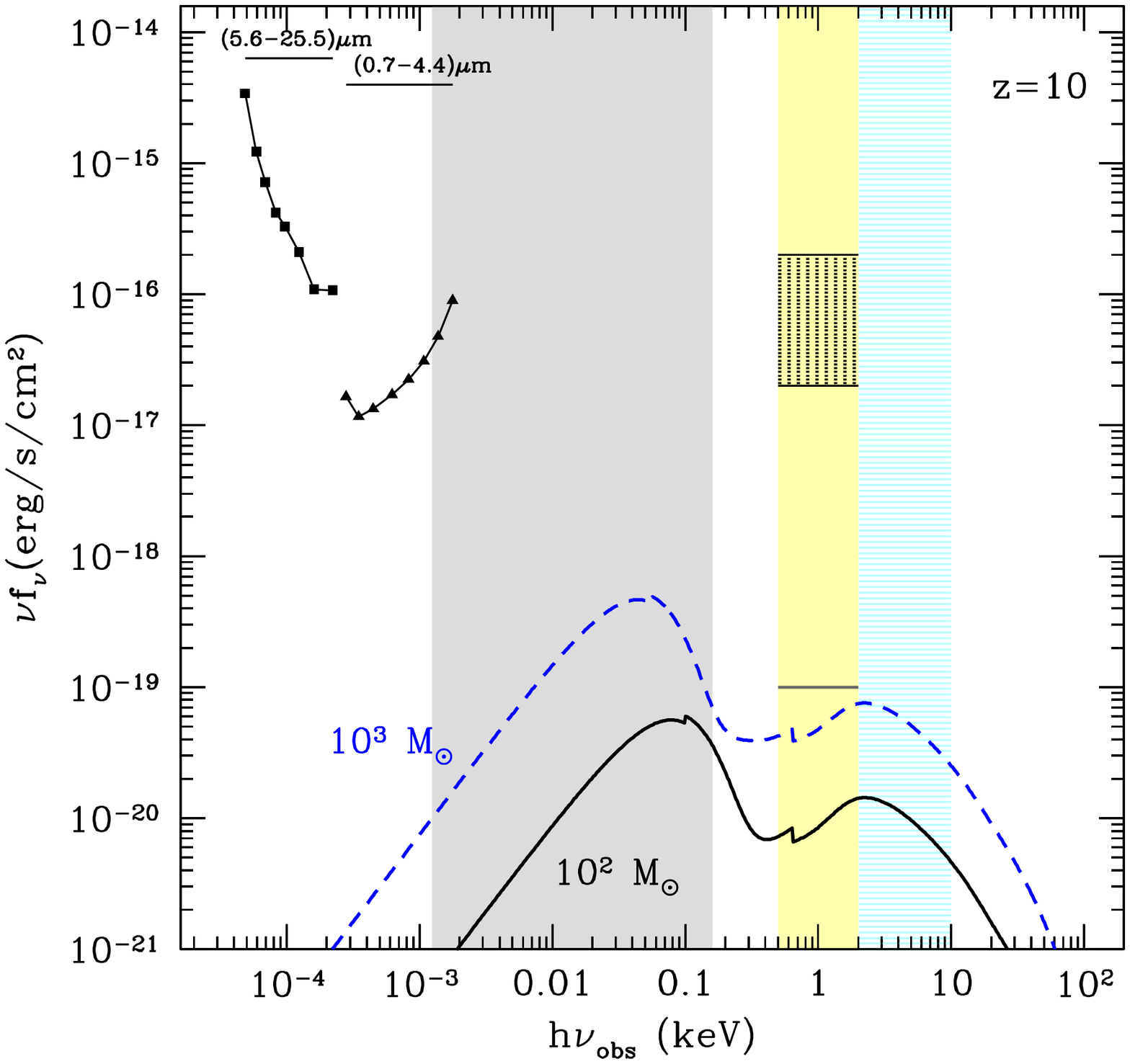}
    \caption{Spectral Energy Distribution (SED) in the observer frame of two light seeds with masses of $100 \msun$ and $1000 \msun$ accreting at the Eddington limit under the most optimistic assumption for detectability, i.e. negligible obscuration and lack of stellar optical/UV emission. We show the SEDs at two different redshifts, $z = 10$ (left panel) and $5$ (right panel). Black lines with points show the sensitivity limits of NIRcam (triangles) and MIRI (squares), on board of {\it JWST}, for a 10 ks exposure. Hard and soft X-ray bands are marked by the cyan and yellow regions. The rectangle in the soft band shows the {\it Athena} area-dependent sensitivity range for the survey designed by \protect\cite{Aird13}.  The horizontal line marks the limiting sensitivity of {\it Lynx} in the soft X-ray for a point source of known position. Finally, the grey shaded area in both panels indicates those wavelengths where emission is expected to be (almost completely) absorbed by the intervening neutral hydrogen along the line-of-sight. 
    }
    \label{fig:emission1}
\end{figure*}

To date, electromagnetic signals from the earliest accreting BHs (seeds) at redshift $z>7.5$
are still missing. Although the Subaru High-$z$ Exploration of Low-Luminosity Quasars  project (SHELLQs) enabled to sample the faint-end tail of the $z\sim 6$ AGN luminosity function, down to a rest-frame ultraviolet absolute magnitude of $M_{1450} = -22$ mag ($L_{\rm bol} \sim 10^{42}$ erg s$^{-1}$ \citealt{Matsuoka18}), no observational signatures of fainter AGN, possibly powered by BHs of $\lesssim {10^7} \msun$ have been found at higher redshift. 

If the high-$z$ population of fainter AGN is powered by heavy, growing seeds, current failed detections might be attributed to the low occupation fraction of this class of BHs, mirroring the rare environmental conditions required for DCBH formation. 
On the other hand, if the growing seed population is dominated by super-Eddington accreting light seeds 
\citep{Inayoshi17, Pezzulli16}, the lack of detection could be due to their short and intermittent activity that is hard to capture within the limited sky-coverage of current surveys \citep{Pezzulli17}.

In addition, X-ray observations of distant, lower mass ($< 10^{5-6} \, \rm M_\odot$) faint AGN are challenging as they may 
be hidden behind the radiation emitted by stellar X-ray binaries forming in the host galaxy, and may suffer from
intrinsic obscuration \citep[e.g.][]{Volonteri17}.

With the next generation of facilities, such as {\it Athena}, early accreting BHs will be within reach, when searched in multi-tiered survey for an observing time of 25 Ms.  The 
maximum redshift, compatible with the limiting sensitivity of the Wide Field Imager is $z \, \leq \, 8$. Observations will provide lower limits on the BH masses, estimated to lie above $10^6\msun$ \citep{Aird13}. 
{\it Lynx}\footnote{https://wwwastro.msfc.nasa.gov/lynx/docs/LynxInterimReport.pdf} is a mission concept to explore the deep 
X-ray Universe, and in long-exposure, multi-tiered surveys it is expected to discover the earliest BHs of $\sim 10^4\msun$ out to $z\sim 10$.

In Fig.~\ref{fig:emission1} we show the spectral energy distribution (SED) of light seeds of $10^2$ (solid lines) and $10^3\, \rm M_\odot$ (dashed lines) at redshift $z=5$ (left panel) and $10$ (right panel). 
In order to set the most favorable conditions for the detectability of unobscured, luminous light seeds, BHs are assumed to grow via gas accretion at the Eddington rate and  the emission from the host galaxy (stellar component) as well as the photoelectric absorption from intervening neutral hydrogen have been neglected.

The SED comprises the optical/UV emission from a (standard) disc multicolor black body spectrum, and the X-ray emission from the hot corona, modeled as a power law with an exponential cut-off at a rest-frame photon energy of 300 keV \citep[see][for details]{Pezzulli17, Valiante18b}. The energy index of the power law in the 2-10 keV interval is correlated with the Eddington ratio ($\lambda_{\rm Ed}$) as $\Gamma=0.23\log \lambda_{\rm Ed}+2.27$ \citep{Brightman2013}.

Modelled fluxes are compared with flux limits of different observatories/missions. Black lines with points show the sensitivity of the {\it JWST} (photometric) instruments NIRcam (triangles, $0.7 - 4.4 \mu \rm m$) and MIRI (squares, $5.6-25.5 \mu \rm m$) for a 10ks exposure. The limiting sensitivity of the concept Lynx, for a point source of known position \footnote{https://wwwastro.msfc.nasa.gov/lynx/docs/science/blackholes.html}, and the Athena area-dependent flux limit range for the survey designed by \cite{Aird13}\footnote{We report the upper and lower flux limits for a $3"$ PSF survey designed as a Wide Field Imager (WFI) wedding cake strategy with single tiers of: $4\times 1$ Ms, $20\times 300$ ks, $75\times 100$ ks and $259\times 10$ ks, for a total collecting area of $2m^2$ at 1 keV and an instrument field of view of $40\times 40$ arcmin} are shown as grey horizontal line and rectangle, respectively.

Fig.~\ref{fig:emission1} shows that light seeds (in this vanilla model) would be too faint to be detectable at $z>5$ with next generation EM facilities. Both the UV flux and the emission from the hot corona are below detectability, even considering the extreme capabilities of {\it Lynx}.

\begin{figure}
    \includegraphics[width=\columnwidth]{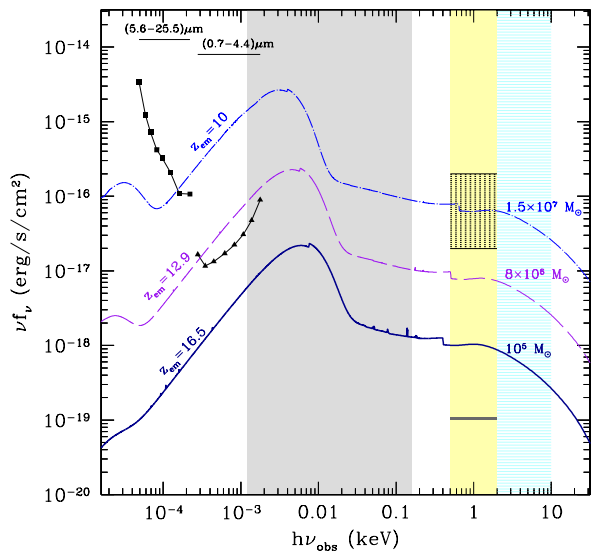}
    \caption{Time-dependent SEDs in the observe frame of a growing heavy seed  forming at $z=16.5$ during the assembly of the $z_{\rm QSO}=6.4$ quasar. The SED of the system is dominated by the emission of the accreting BH.  Starting from an initial mass of $10^5 \msun$, the emission of a heavy seed is shown at three different ages, labelled with their emission redshift. At $z=10$  the BH mass is $1.5\times 10^7 \msun$. The shaded areas, and the {\it JWST, Athena}, and {\it Lynx} limiting sensitivities are  indicated as in Fig.~\ref{fig:emission1}. 
    The gray shaded region here shows the wavelength range affected by absorption along the light of sight for the source at $z=10$.
    }
    \label{fig:emission2}
\end{figure}

Following \citet{Valiante18b}, we also show in Fig. \ref{fig:emission2} the evolving SED of a heavy seed that forms at $z=16.5$ among the progenitors of the $z_{\rm QSO}=6.4$ quasar, and that grows via gas accretion only. 
Starting from an initial mass of $10^5 \, \rm M_\odot$, the seed experiences Eddington-limited growth during the
250 Myr of "isolated" evolution of the system \citep[i.e. before a galaxy merger occurs][]{Valiante18b}. 
In this case, both the stellar and accreting BH intrinsic emission have been reprocessed through the host galaxy ISM, combining the \textsc{GQd} model predictions (galaxy SFR, BH accretion rate, ISM metallicity and dust-to-gas ratio, etc.) with the radiative transfer code Cloudy (\citealt{Ferland13}, see \citealt{Valiante18b} for details.).

The SED of the growing heavy seed is shown at three different ages (labelled with their emission redshift), ending at
$z = 10$ when the BH mass is $\sim 1.5\times 10^7 \, \rm M_\odot$. Although we include the starburst contribution, in this case the emission is completely dominated by the accreting BH at all redshifts, in other words the AGN is way more luminous then the host galaxy stellar component and it is potentially detectable by both {\it Athena} and NIRcam on board {\it JWST} \citep{Valiante18b}.

\section{Discussion}\label{sec:discussion}
If the growth of seeds is regulated by gas accretion in halos experiencing multiple mergers, their modeling encompasses a rich and complex variety of physical processes. During galaxy assembly ruled by mergers and gas inflows from the cosmic web, the formation of binary seeds appears highly probable if not inevitable. 

In this work we used the semi-analytical, data-constrained, hierarchical model \textsc{GQd} \citep[][]{Valiante14, Valiante16, Valiante18a} to track the formation of SMBHs starting from the first stars and first BH seeds, light and heavy, following the formation of the earliest BH binaries and their coalescence driven by triple interactions.
We assumed that BBHs form within at most few Myr in $100\%$ of major halo-halo mergers and that a triplet form in $100\%$ of triple/multiple BH encounters (see Section~\ref{sec:BBHdynamics}).
The sinking timescale of BHs on kpc-to-pc scales is usually set by dynamical friction against background stars and gas. 
The halo mass ratio, BH intrinsic masses (customarily in excess of $10^6\msun$), DM profiles, redshift-dependent gas fraction and galaxy morphology, presence of irregular substructures and even the spatial and mass resolution of simulations all control the formation/failure of a bound system \citep[see e.g.][and references therein]{Callegari09,  Fiacconi13, Capelo2015, Tamburello2017,  Pfister2017,Tamfal2018}.
In zoomed-in high redshift ($z\sim 9$) simulations of dwarf proto-galaxies, dynamical friction against stars is found to be the main process of BH orbital decay for  $\sim 10^5 \msun$ seeds, while erratic dynamics is seen below this mass, implying either rapid decay or BH wandering/ejection and the presence of  multiple BHs in a galaxy, each inherited from a different merger \citep{Pfister2019}.  
Interestingly, at high redshifts ($z>6$) and for BHs of $\sim 10^6\msun$, global or  bar-induced torques in some cases appear to be more efficient than dynamical friction in promoting BH binary formation on timescales comparable to the local Hubble time at those redshift \citep{Bortolas2020}. 
Moreover, additional kpc-scale delays can further alter the above picture \citep[see e.g.][]{Barausse2020}
Yet, the process of light seeds binaries formation/merger is unexplored in cosmological simulations, as capturing their dynamics requires extreme high spatial and mass resolution.

In our approach cosmologically-driven BH mergers are triggered only via triplet formation following the prescriptions of \citet[][]{Bonetti18a,Bonetti18b} with the sole difference that we approximate the triple interaction as instantaneous, neglecting the triplet lifetime. This relies on the fact that, although the triplet lifetime shows a log-normal distribution with a mean value of $\sim 250$ Myr, this is mostly due to the dynamical friction phase, which we do not model here\footnote{It should also be noted that  the stellar environment of Bonetti et al. simulations was calibrated against low-$z$ galaxies, and as such are not directly applicable to the problem at hand. First, due to the shorter local dynamical time, one might expect a much faster evolution in dense protogalaxies at high redshift. Second, the evolution might well be dominated by dynamical friction against the dense gaseous background rather than stars.}. Once the three-body interaction becomes effective, the associated timescale to resolve the triplet (either a merger or an ejection) is actually much shorter ($\sim$ few Myr), justifying our assumption of instantaneous interaction. 
Triplets also have a limited efficiency (at most $\sim 30\%$) in triggering BH mergers and we expect that a large fraction of triple encounters end up with a "stalled" left-over system \citep[][]{Bonetti18b}.

Neglecting the physical delays could imply a higher fraction of mergers at earlier times/at lower mass ranges. Depending on the delay time \citep[i.e. time spent by the triplet before coalescence, as computed by][]{Bonetti18a} the merger event could be shifted at lower redshifts and, in the meantime, the inner binary could grow in mass via gas accretion onto the two components (thus changing the merger probability). In addition, when the dynamical merger timescale is longer than our binary lifetime (defined in Section 4.1), we may expect an additional intruder to interact with the triplet, further complicating the scenario (and the description of dynamical processes). We plan to study these more complex aspects in a future work.

{In our implementation, we also neglected the effect of stellar hardening and viscous migration in driving the two BHs down to the GW-driven domain. It should be noted that, considering additional hardening timescales due to binary-gas disc interactions and/or stellar-dominated processes may contribute to the population of merging BHs \citep[e.g.][and references therein]{Bortolas16, Bortolas18a, Biava19,ArcaSedda19,Lima2020}. Therefore, in this respect our results should be viewed as conservative and, in a forthcoming work, we aim at introducing an improved description of more realistic BH dynamics and merger timescales to analyze their impact on SMBHs growth and BH merger history.
We expect efficient stellar/gas hardening to have a major impact on the "stalled" left-over binaries (i.e. in the case in which triple interactions fail in triggering BH coalescence) and/or when the triplet-driven mergers require long timescales \citep[>1 Gyr, as e.g. following the ejection of one BH][]{Bonetti18b}. In environments in which stellar/gas driven shrinking proceeds on relatively short timescales ($<100-300\, \rm {Myr}$) the binary may be efficiently driven down to the GW emission phase \citep[][]{Bortolas18b, ArcaSedda19} even before a triplet forms thus, affecting the number and redshift of the mergers.
}

Full control of the BH dynamics down to the GW driven domain is fundamental when predicting the rate of BH coalescences alongside the hierarchical assembly of galaxies. 
This has been investigated in a number of studies so far, under different assumptions and approaches regarding the merger timescales \citep[e.g.][and references therein]{Enoki2005, SesanaGair11, Klein2016, Tamanini16, Ricarte2018, Bonetti19, Dayal2019,Katz2020, Volonteri2020}. 

In our analysis we simulate the histories of SMBHs and their host galaxy, forming in rare, highly biased regions of the Universe. Thus, a direct comparison of our results with the studies mentioned above is difficult, as these usually describe populations of galaxies/AGN in an "average" region of the Universe.

Using the SAM {\it Delphi}, \citet[][]{Dayal2019} find that binaries with total masses of $10^{3.5}-10^5 \msun$ are detectable, with a $\rm S/N > 7$, in the redshift range $z\sim 5-13$, with the large fraction being mergers of light seeds (called "Type 1" mergers). This is consistent with our predictions shown in Fig.~\ref{fig:WFplots}. 

Within a zoomed-in, re-simulated, region of ($15 \, h^{-1}$ Mpc)$^3$ extracted from the BlueTides cosmological hydrodynamic simulation, \citet[][]{Huang2019} examined the early growth of $z>6$ SMBHs, running different sets of simulations for three different BH seed masses: $5\times 10^3$, $5\times 10^4$ and $5\times 10^5 \, h^{-1} \msun$. All seed scenarios eventually converge to form SMBH of $\sim 10^9 \msun$ provided that the halo mass threshold to BH seed mass ratio is the same (constant). In their simulations the rate/number of BH mergers is higher in the low-mass seed scenario (8 mergers), as lighter seeds are more common/abundant than the more massive ones. Four of such mergers occur at $z>12$ with total masses of $10^4-10^6 \msun$, thus being potentially detectable with LISA. This result is consistent with our findings in the $z_{\rm QSO}=6.4$ merger history. In contrast, as a consequence of the different BH seeding and dynamics prescriptions, their massive seed model ($\sim 5\times 10^5 \msun$) does not predict any merger until $z<6$.

\section{Conclusions}\label{sec:conclusions}
Our model suggests that a statistical inference of the mass distribution and relative occurrence of the earliest BH mergers, if/when provided by the combination of ET and LISA detections, will offer a unique insight into the earliest BH seeds formation epoch and its evolution across cosmic time. 
On the other hand, thanks to the better sensitivity of CE at higher frequencies and thus lower BH stellar masses, CE observations will be fundamental to study the complementary population of stellar-mass BBHs with mass $\leq 100\msun$ out to $z\sim 10-15$.

As commonly expected, an observational signature of the light BH seed channel could be the higher occupation fraction and thus a higher merger rate compared to the heavy seed one \citep[e.g.][]{Sesana07, SesanaGair11, Klein2016, Ricarte2018, Bonetti19, Huang2019}. X-rays, deep field, observations may help in discriminating the imprints of different BH seeds (e.g. in the high-$z$ luminosity functions), although it will be challenging to uniquely disentangle their EM observational features \citep[][]{Pacucci15, Natarajan17, Volonteri17, Valiante18b, Ricarte2018}. 

\begin{figure*}
    \centering
    \includegraphics[width=0.8\textwidth]{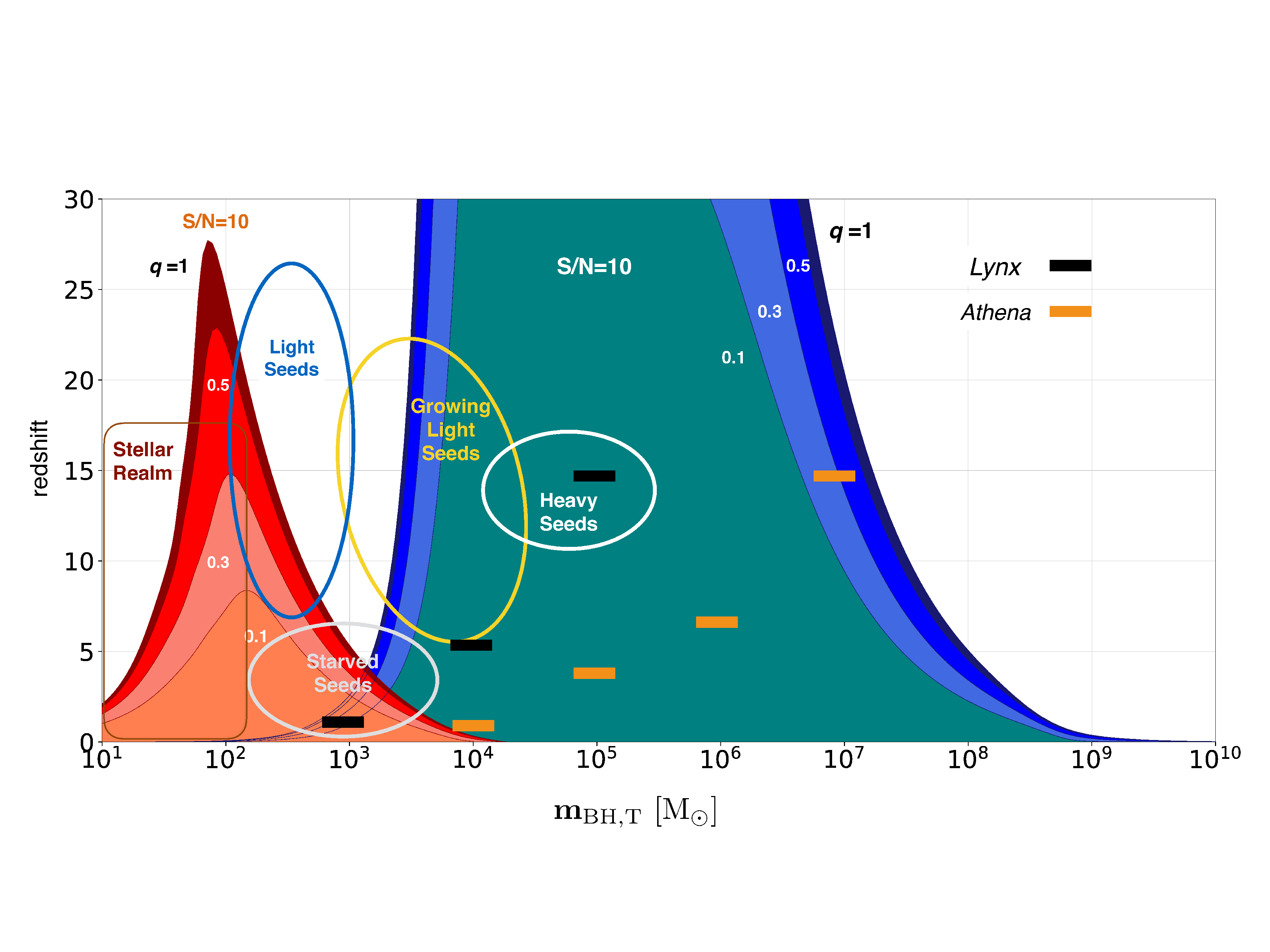}
    \caption{ 
    The GW and EM landscape. Color-coded areas give the average GW horizon computed for a detection threshold equal to $\rm S/N=10$: contour lines refer to binaries with mass ratios $q=1,0.5,0.3,0.1$ both in the ET and LISA bandwidth. Upper limits (shown as thick horizontal bars) indicate the sensitivity of the deepest pointing, in the $[0.5-2]$ keV observed band, by {\it Athena} (orange) and {\it Lynx} (black) given the limiting fluxes of $2.4\times 10^{-17}$ and $10^{-19}$ erg s$^{-1}$ cm$^{-2}$, respectively. The upper limits are inferred assuming that BHs are emitting at the Eddington limit and adopting a bolometric correction ($L_{\rm X}/L_{\rm bol})$ of $10\%$. Ellipses highlight the islands in the $z-m_{\rm BH,T}$ plane where light (blue) and heavy (white) seeds are expected to form as well as where light seeds are expected to grow via accretion and mergers (yellow). The transit to the SMBH domain covers the entire LISA area and EM observations are key to discover the high-mass tail of the SMBH distribution. The light-grey ellipse below $z\sim 5$ marks the population of long-living "starved" seeds. Note that in this island, coordinated multi-band observations are possible having LISA the capability to first follow the early inspiral in intermediate-mass black holes and ET the merger phase, enhancing the ability to carry on precise measurements of the source parameters also at $z\sim 5$ \citep{Jani2019}.
    The islands have overlap with the GW horizon, but an empty inaccessible region is present between ET and LISA, corresponding to the Deci-Hz GW domain. The island corresponding to the stellar realm is included, on the left, for comparison.
    }
    \label{fig:GWsummary}
\end{figure*}

Detecting the GW signals of BHs of $\sim 100\msun$ up to $\sim  10^7\msun$ from cosmic dawn to the present will enable us to unveil if seeds are the {\it fil rouge} connecting the stellar BHs to the supermassive BHs, or if a desert and genetic division exists between the two populations \citep{Colpi2019-book}.

Figure~\ref{fig:GWsummary} summarizes the limiting GW and EM sensitivities in the $m_{\rm BH,T} - z$ plane. Waterfall plots for LISA (blue) and ET (red) for a signal-to-noise ratio S/N=10 are reported as a function of the merging BH binary mass ratio $q=1, 0.5, 0.3, 0.1$ while upper limits show the highest redshift at which an accreting BH of given mass (equivalent to the mass of a BBH) is detectable by {\it Athena} (orange), at the deepest survey layer limiting flux of $2.4\times 10^{-17}$ erg s$^{-1}$ cm$^{-2}$, and by {\it Lynx} (black) at the limiting sensitivity level of $10^{-19}$ erg s$^{-1}$ cm$^{-2}$. These upper limits are computed assuming that accreting BHs emit at the Eddington luminosity, $L=L_{\rm Edd}$, with 10\% of the flux emerging in the hard X-ray bandwidth, suitably redshifted.\footnote{A Hard-to-Soft X-ray luminosity conversion factor of 1.35 is taken into account, for a power-law spectrum with photon index $\Gamma = 1.9$.} 
The ellipses drawn in the figure mark the different regions where light (blue ellipse) and heavy (white ellipse) seeds are expected to form and where growing (yellow ellipse) and "starved" (gray ellipse) light seeds are expected to reside. For comparison, on the left of the figure we plot also the region corresponding to stellar-mass BHs, under the assumptions that these come from Pop~II stellar binaries formed in the field and in higher metallicity environments, and that their total masses can extend up to few $\sim 100 \msun$, with a potential superposition with our "starved" seed population.

ET with sensitivity down to a few Hz shall have the unique capability of discovering the earliest BH binaries in the range of stellar  BHs, light and medium-weight seeds forming in the Universe, probing the existence of these rare transitional objects that happen to evolve into SMBHs through gas accretion and mergers under favorable cosmic conditions. 
{\it ET will be the only instrument that will let us discover light BH seeds forming at cosmic dawn}. 

On the other hand, if these seeds fail to grow, they may be present in galaxies at lower redshift.  3G detectors shall have the sensitivity to reveal such failed seeds, that we define as  "starved" seeds. 
{\it Discovering BHs in this uncharted territory will be groundbreaking.}

As light/medium-weight seeds evolve via accretion and mergers, they will transit across the LISA bandwidth and the match between ET and LISA events will statistically shed light into the seeding mechanism. LISA has also the potential to detect the rare heavy seeds in their transit to become supermassive. The lack of events on the right side of ET waterfall plot could be an indication that only heavy seeds are the progenitor of the SMBHs or that light seeds grow at a very fast (super-Eddington) rate, following their formation without experiencing cosmologically driven mergers.

Finally, there are planned experimental programs employing atom interferometers, like the Atom Interferometer Observatory and Network \citep[AION,][]{Badurina2020} which propose
to explore GWs in the mid-frequency range, filling the gap between CE/ET and LISA.

Deep EM observations of galaxies and active BHs at redshifts $z\sim 8-10$ with forthcoming and next-generation facilities combined with independent observations of coalescing BHs with GW observatories will offer the first ever view of the young Universe, by capturing the first moment of star and BH formation in the earliest galaxies. While {\it JWST, Athena} and {\it Lynx} (if in operation) will see little patches of the deep Universe to unveil the dawn of galaxies and accreting black holes, ET and LISA will witness the dawn of black hole binaries. 

In a companion paper we will investigate in detail the expected accuracy of parameter recovery from gravitational wave signals from light seeds observed in CE/ET and from growing light seeds and heavy seeds in LISA.   As discussed in Section \ref{sec:QSOs}, the waveform in the CE/ET sensitive band will comprise only a few cycles and consequently accurate recovery of parameters will be challenging.  We will investigate whether, with accurate waveforms incorporating spin effects and higher harmonics, we will be enable differentiation of candidate light seeds from black hole mergers of stellar origin. We will also carry on parameter estimation analysis of  the high redshift seeds detectable with LISA during their slow, adiabatic inspiral. 

As discussed in Section~\ref{sec:discussion}, including the physics of BH dynamics (e.g. realistic astrophysical time delays) is critical for any reliable characterization of the merging BBH populations across the cosmic history (as well as for the evaluation of the merger rates). We aim to improve the model presented here including proper binaries formation/merger timescales and extending our investigations to {\it (i)} additional seed flavours (e.g. including the medium-weight channel and a mass function for heavy seeds) and to {\it (ii)} wider ranges of DM halos masses and redshift (e.g. to quantify BBHs occupation fraction and LISA/ET/CE merger rates across cosmic epochs).

\section*{Acknowledgements}
We thank the anonymous referee for the constructive comments, Evan Hall for sharing agreement on the ET waterfall plots, and Bangalore Sathyaprakash for enlightening discussions. RV thanks Luca Zappacosta for productive discussions and suggestions. RV, RS and MC acknowledge support from the Amaldi Research Center
funded by the MIUR program {\it Dipartimento di Eccellenza}
(CUP:B81I18001170001). MB, MC, FH, AM and RV acknowledge the networking support by the COST Action CA16104 and funding from the INFN  TEONGRAV  specific  initiative, and MIUR under the grant PRIN 2017-MB8AEZ. 
SF and CM acknowledge support from the Science and Technology Facilities Council (STFC) grant ST/L000962/1, and European Research Council Consolidator Grant 647839.

\section*{Data Availability}
The simulated data underlying this article will be shared on reasonable request to the corresponding author.


\bibliographystyle{mnras}
\bibliography{references} 

\begin{thebibliography}{}
\makeatletter
\relax
\def\mn@urlcharsother{\let\do\@makeother \do\$\do\&\do\#\do\^\do\_\do\%\do\~}
\def\mn@doi{\begingroup\mn@urlcharsother \@ifnextchar [ {\mn@doi@}
  {\mn@doi@[]}}
\def\mn@doi@[#1]#2{\def\@tempa{#1}\ifx\@tempa\@empty \href
  {http://dx.doi.org/#2} {doi:#2}\else \href {http://dx.doi.org/#2} {#1}\fi
  \endgroup}
\def\mn@eprint#1#2{\mn@eprint@#1:#2::\@nil}
\def\mn@eprint@arXiv#1{\href {http://arxiv.org/abs/#1} {{\tt arXiv:#1}}}
\def\mn@eprint@dblp#1{\href {http://dblp.uni-trier.de/rec/bibtex/#1.xml}
  {dblp:#1}}
\def\mn@eprint@#1:#2:#3:#4\@nil{\def\@tempa {#1}\def\@tempb {#2}\def\@tempc
  {#3}\ifx \@tempc \@empty \let \@tempc \@tempb \let \@tempb \@tempa \fi \ifx
  \@tempb \@empty \def\@tempb {arXiv}\fi \@ifundefined
  {mn@eprint@\@tempb}{\@tempb:\@tempc}{\expandafter \expandafter \csname
  mn@eprint@\@tempb\endcsname \expandafter{\@tempc}}}

\bibitem[\protect\citeauthoryear{{Abbott} et~al.,}{{Abbott}
  et~al.}{2017}]{CE17}
{Abbott} B.~P.,  et~al., 2017, \mn@doi [Classical and Quantum Gravity]
  {10.1088/1361-6382/aa51f4}, \href
  {https://ui.adsabs.harvard.edu/abs/2017CQGra..34d4001A} {34, 044001}

\bibitem[\protect\citeauthoryear{{Abel}, {Bryan}  \& {Norman}}{{Abel}
  et~al.}{2002}]{Abel02}
{Abel} T.,  {Bryan} G.~L.,   {Norman} M.~L.,  2002, \mn@doi [Science]
  {10.1126/science.295.5552.93}, \href
  {http://adsabs.harvard.edu/abs/2002Sci...295...93A} {295, 93}

\bibitem[\protect\citeauthoryear{{Agarwal}, {Khochfar}, {Johnson}, {Neistein},
  {Dalla Vecchia}  \& {Livio}}{{Agarwal} et~al.}{2012}]{Agarwal12}
{Agarwal} B.,  {Khochfar} S.,  {Johnson} J.~L.,  {Neistein} E.,  {Dalla
  Vecchia} C.,   {Livio} M.,  2012, \mn@doi [\mnras]
  {10.1111/j.1365-2966.2012.21651.x}, \href
  {http://adsabs.harvard.edu/abs/2012MNRAS.425.2854A} {425, 2854}

\bibitem[\protect\citeauthoryear{{Aird}, {Comastri}, {Brusa}, {Cappelluti},
  {Moretti}, {Vanzella}  et~al.}{{Aird} et~al.}{2013}]{Aird13}
{Aird} J.,  {Comastri} A.,  {Brusa} M.,  {Cappelluti} N.,  {Moretti} A.,
  {Vanzella} E.,   et~al., 2013, preprint, \href
  {http://adsabs.harvard.edu/abs/2013arXiv1306.2325A} {} (\mn@eprint {arXiv}
  {1306.2325})

\bibitem[\protect\citeauthoryear{{Alexander} \& {Natarajan}}{{Alexander} \&
  {Natarajan}}{2014}]{Alexander2014}
{Alexander} T.,  {Natarajan} P.,  2014, \mn@doi [Science]
  {10.1126/science.1251053}, \href
  {https://ui.adsabs.harvard.edu/abs/2014Sci...345.1330A} {345, 1330}

\bibitem[\protect\citeauthoryear{{Amaro-Seoane} et~al.}{{Amaro-Seoane}
  et~al.}{2017}]{LISA17}
{Amaro-Seoane} P.,  et~al., 2017, preprint, \href
  {http://adsabs.harvard.edu/abs/2017arXiv170200786A} {} (\mn@eprint {arXiv}
  {1702.00786})

\bibitem[\protect\citeauthoryear{{Arca Sedda} et~al.,}{{Arca Sedda}
  et~al.}{2019a}]{Voyage2019}
{Arca Sedda} M.,  et~al., 2019a, arXiv e-prints, \href
  {https://ui.adsabs.harvard.edu/abs/2019arXiv190811375A} {p. arXiv:1908.11375}

\bibitem[\protect\citeauthoryear{{Arca Sedda}, {Berczik}, {Capuzzo-Dolcetta},
  {Fragione}, {Sobolenko}  \& {Spurzem}}{{Arca Sedda}
  et~al.}{2019b}]{ArcaSedda19}
{Arca Sedda} M.,  {Berczik} P.,  {Capuzzo-Dolcetta} R.,  {Fragione} G.,
  {Sobolenko} M.,   {Spurzem} R.,  2019b, \mn@doi [\mnras]
  {10.1093/mnras/sty3458}, \href
  {https://ui.adsabs.harvard.edu/abs/2019MNRAS.484..520A} {484, 520}

\bibitem[\protect\citeauthoryear{{Askar}, {Szkudlarek}, {Gondek-Rosi{\'n}ska},
  {Giersz}  \& {Bulik}}{{Askar} et~al.}{2017}]{Askar2017}
{Askar} A.,  {Szkudlarek} M.,  {Gondek-Rosi{\'n}ska} D.,  {Giersz} M.,
  {Bulik} T.,  2017, \mn@doi [\mnras] {10.1093/mnrasl/slw177}, \href
  {https://ui.adsabs.harvard.edu/abs/2017MNRAS.464L..36A} {464, L36}

\bibitem[\protect\citeauthoryear{{Ba{\~n}ados} et~al.,}{{Ba{\~n}ados}
  et~al.}{2016}]{Banados16}
{Ba{\~n}ados} E.,  et~al., 2016, \mn@doi [\apjs] {10.3847/0067-0049/227/1/11},
  \href {http://adsabs.harvard.edu/abs/2016ApJS..227...11B} {227, 11}

\bibitem[\protect\citeauthoryear{{Ba{\~n}ados} et~al.,}{{Ba{\~n}ados}
  et~al.}{2018}]{Banados18}
{Ba{\~n}ados} E.,  et~al., 2018, \mn@doi [\nat] {10.1038/nature25180}, \href
  {http://adsabs.harvard.edu/abs/2018Natur.553..473B} {553, 473}

\bibitem[\protect\citeauthoryear{{Badurina} et~al.,}{{Badurina}
  et~al.}{2020}]{Badurina2020}
{Badurina} L.,  et~al., 2020, \mn@doi [\jcap] {10.1088/1475-7516/2020/05/011},
  \href {https://ui.adsabs.harvard.edu/abs/2020JCAP...05..011B} {2020, 011}

\bibitem[\protect\citeauthoryear{{Baldassare}, {Reines}, {Gallo}  \&
  {Greene}}{{Baldassare} et~al.}{2015}]{Baldassare15}
{Baldassare} V.~F.,  {Reines} A.~E.,  {Gallo} E.,   {Greene} J.~E.,  2015,
  \mn@doi [\apjl] {10.1088/2041-8205/809/1/L14}, \href
  {http://adsabs.harvard.edu/abs/2015ApJ...809L..14B} {809, L14}

\bibitem[\protect\citeauthoryear{{Barausse}}{{Barausse}}{2012}]{Barausse2012}
{Barausse} E.,  2012, \mn@doi [\mnras] {10.1111/j.1365-2966.2012.21057.x},
  \href {https://ui.adsabs.harvard.edu/abs/2012MNRAS.423.2533B} {423, 2533}

\bibitem[\protect\citeauthoryear{{Barausse}, {Morozova}  \&
  {Rezzolla}}{{Barausse} et~al.}{2012}]{Barausse12}
{Barausse} E.,  {Morozova} V.,   {Rezzolla} L.,  2012, \mn@doi [\apj]
  {10.1088/0004-637X/758/1/63}, \href
  {http://adsabs.harvard.edu/abs/2012ApJ...758...63B} {758, 63}

\bibitem[\protect\citeauthoryear{{Barausse}, {Dvorkin}, {Tremmel}, {Volonteri}
  \& {Bonetti}}{{Barausse} et~al.}{2020}]{Barausse2020}
{Barausse} E.,  {Dvorkin} I.,  {Tremmel} M.,  {Volonteri} M.,   {Bonetti} M.,
  2020, arXiv e-prints, \href
  {https://ui.adsabs.harvard.edu/abs/2020arXiv200603065B} {p. arXiv:2006.03065}

\bibitem[\protect\citeauthoryear{{Begelman}, {Blandford}  \& {Rees}}{{Begelman}
  et~al.}{1980}]{Begelman80}
{Begelman} M.~C.,  {Blandford} R.~D.,   {Rees} M.~J.,  1980, \mn@doi [\nat]
  {10.1038/287307a0}, \href {http://adsabs.harvard.edu/abs/1980Natur.287..307B}
  {287, 307}

\bibitem[\protect\citeauthoryear{{Begelman}, {Volonteri}  \& {Rees}}{{Begelman}
  et~al.}{2006}]{Begelman06}
{Begelman} M.~C.,  {Volonteri} M.,   {Rees} M.~J.,  2006, \mn@doi [\mnras]
  {10.1111/j.1365-2966.2006.10467.x}, \href
  {http://adsabs.harvard.edu/abs/2006MNRAS.370..289B} {370, 289}

\bibitem[\protect\citeauthoryear{{Biava}, {Colpi}, {Capelo}, {Bonetti},
  {Volonteri}, {Tamfal}, {Mayer}  \& {Sesana}}{{Biava} et~al.}{2019}]{Biava19}
{Biava} N.,  {Colpi} M.,  {Capelo} P.~R.,  {Bonetti} M.,  {Volonteri} M.,
  {Tamfal} T.,  {Mayer} L.,   {Sesana} A.,  2019, \mn@doi [\mnras]
  {10.1093/mnras/stz1614}, \href
  {https://ui.adsabs.harvard.edu/abs/2019MNRAS.487.4985B} {487, 4985}

\bibitem[\protect\citeauthoryear{{Bischetti} et~al.,}{{Bischetti}
  et~al.}{2019}]{Bischetti19}
{Bischetti} M.,  et~al., 2019, \mn@doi [\aap] {10.1051/0004-6361/201935524},
  \href {https://ui.adsabs.harvard.edu/abs/2019A&A...628A.118B} {628, A118}

\bibitem[\protect\citeauthoryear{{Bonetti}, {Haardt}, {Sesana}  \&
  {Barausse}}{{Bonetti} et~al.}{2016}]{Bonetti16}
{Bonetti} M.,  {Haardt} F.,  {Sesana} A.,   {Barausse} E.,  2016, \mn@doi
  [\mnras] {10.1093/mnras/stw1590}, \href
  {https://ui.adsabs.harvard.edu/abs/2016MNRAS.461.4419B} {461, 4419}

\bibitem[\protect\citeauthoryear{{Bonetti}, {Sesana}, {Barausse}  \&
  {Haardt}}{{Bonetti} et~al.}{2018a}]{Bonetti18b}
{Bonetti} M.,  {Sesana} A.,  {Barausse} E.,   {Haardt} F.,  2018a, \mn@doi
  [\mnras] {10.1093/mnras/sty874}, \href
  {https://ui.adsabs.harvard.edu/abs/2018MNRAS.477.2599B} {477, 2599}

\bibitem[\protect\citeauthoryear{{Bonetti}, {Haardt}, {Sesana}  \&
  {Barausse}}{{Bonetti} et~al.}{2018b}]{Bonetti18a}
{Bonetti} M.,  {Haardt} F.,  {Sesana} A.,   {Barausse} E.,  2018b, \mn@doi
  [\mnras] {10.1093/mnras/sty896}, \href
  {https://ui.adsabs.harvard.edu/abs/2018MNRAS.477.3910B} {477, 3910}

\bibitem[\protect\citeauthoryear{{Bonetti}, {Sesana}, {Haardt}, {Barausse}  \&
  {Colpi}}{{Bonetti} et~al.}{2019}]{Bonetti19}
{Bonetti} M.,  {Sesana} A.,  {Haardt} F.,  {Barausse} E.,   {Colpi} M.,  2019,
  \mn@doi [\mnras] {10.1093/mnras/stz903}, \href
  {https://ui.adsabs.harvard.edu/abs/2019MNRAS.486.4044B} {486, 4044}

\bibitem[\protect\citeauthoryear{{Booth} \& {Schaye}}{{Booth} \&
  {Schaye}}{2009}]{BoothSchaye09}
{Booth} C.~M.,  {Schaye} J.,  2009, \mn@doi [\mnras]
  {10.1111/j.1365-2966.2009.15043.x}, \href
  {http://adsabs.harvard.edu/abs/2009MNRAS.398...53B} {398, 53}

\bibitem[\protect\citeauthoryear{{Bortolas}, {Gualandris}, {Dotti}, {Spera}  \&
  {Mapelli}}{{Bortolas} et~al.}{2016}]{Bortolas16}
{Bortolas} E.,  {Gualandris} A.,  {Dotti} M.,  {Spera} M.,   {Mapelli} M.,
  2016, \mn@doi [\mnras] {10.1093/mnras/stw1372}, \href
  {https://ui.adsabs.harvard.edu/abs/2016MNRAS.461.1023B} {461, 1023}

\bibitem[\protect\citeauthoryear{{Bortolas}, {Mapelli}  \& {Spera}}{{Bortolas}
  et~al.}{2018a}]{Bortolas18a}
{Bortolas} E.,  {Mapelli} M.,   {Spera} M.,  2018a, \mn@doi [\mnras]
  {10.1093/mnras/stx2795}, \href
  {https://ui.adsabs.harvard.edu/abs/2018MNRAS.474.1054B} {474, 1054}

\bibitem[\protect\citeauthoryear{{Bortolas}, {Gualandris}, {Dotti}  \&
  {Read}}{{Bortolas} et~al.}{2018b}]{Bortolas18b}
{Bortolas} E.,  {Gualandris} A.,  {Dotti} M.,   {Read} J.~I.,  2018b, \mn@doi
  [\mnras] {10.1093/mnras/sty775}, \href
  {https://ui.adsabs.harvard.edu/abs/2018MNRAS.477.2310B} {477, 2310}

\bibitem[\protect\citeauthoryear{{Bortolas}, {Capelo}, {Zana}, {Mayer},
  {Bonetti}, {Dotti}, {Davies}  \& {Madau}}{{Bortolas}
  et~al.}{2020}]{Bortolas2020}
{Bortolas} E.,  {Capelo} P.~R.,  {Zana} T.,  {Mayer} L.,  {Bonetti} M.,
  {Dotti} M.,  {Davies} M.~B.,   {Madau} P.,  2020, arXiv e-prints, \href
  {https://ui.adsabs.harvard.edu/abs/2020arXiv200502409B} {p. arXiv:2005.02409}

\bibitem[\protect\citeauthoryear{{Brightman} et~al.,}{{Brightman}
  et~al.}{2013}]{Brightman2013}
{Brightman} M.,  et~al., 2013, \mn@doi [\mnras] {10.1093/mnras/stt920}, \href
  {https://ui.adsabs.harvard.edu/abs/2013MNRAS.433.2485B} {433, 2485}

\bibitem[\protect\citeauthoryear{{Bromm} \& {Loeb}}{{Bromm} \&
  {Loeb}}{2004}]{BrommLoeb04}
{Bromm} V.,  {Loeb} A.,  2004, \mn@doi [\na] {10.1016/j.newast.2003.12.006},
  \href {http://adsabs.harvard.edu/abs/2004NewA....9..353B} {9, 353}

\bibitem[\protect\citeauthoryear{{Callegari}, {Mayer}, {Kazantzidis}, {Colpi},
  {Governato}, {Quinn}  \& {Wadsley}}{{Callegari} et~al.}{2009}]{Callegari09}
{Callegari} S.,  {Mayer} L.,  {Kazantzidis} S.,  {Colpi} M.,  {Governato} F.,
  {Quinn} T.,   {Wadsley} J.,  2009, \mn@doi [\apjl]
  {10.1088/0004-637X/696/1/L89}, \href
  {http://adsabs.harvard.edu/abs/2009ApJ...696L..89C} {696, L89}

\bibitem[\protect\citeauthoryear{{Capelo}, {Volonteri}, {Dotti}, {Bellovary},
  {Mayer}  \& {Governato}}{{Capelo} et~al.}{2015}]{Capelo2015}
{Capelo} P.~R.,  {Volonteri} M.,  {Dotti} M.,  {Bellovary} J.~M.,  {Mayer} L.,
   {Governato} F.,  2015, \mn@doi [\mnras] {10.1093/mnras/stu2500}, \href
  {https://ui.adsabs.harvard.edu/abs/2015MNRAS.447.2123C} {447, 2123}

\bibitem[\protect\citeauthoryear{{Chon} \& {Omukai}}{{Chon} \&
  {Omukai}}{2020}]{Chon20}
{Chon} S.,  {Omukai} K.,  2020, arXiv e-prints, \href
  {https://ui.adsabs.harvard.edu/abs/2020arXiv200106491C} {p. arXiv:2001.06491}

\bibitem[\protect\citeauthoryear{{Chon}, {Hirano}, {Hosokawa}  \&
  {Yoshida}}{{Chon} et~al.}{2016}]{Chon16}
{Chon} S.,  {Hirano} S.,  {Hosokawa} T.,   {Yoshida} N.,  2016, \mn@doi [\apj]
  {10.3847/0004-637X/832/2/134}, \href
  {http://adsabs.harvard.edu/abs/2016ApJ...832..134C} {832, 134}

\bibitem[\protect\citeauthoryear{{Colpi}}{{Colpi}}{2014}]{Colpi14}
{Colpi} M.,  2014, \mn@doi [\ssr] {10.1007/s11214-014-0067-1}, \href
  {http://adsabs.harvard.edu/abs/2014SSRv..183..189C} {183, 189}

\bibitem[\protect\citeauthoryear{{Colpi}}{{Colpi}}{2019}]{Colpi2019-book}
{Colpi} M.,  2019, {Probing the formation of the seeds of supermassive black
  holes with gravitational waves}.
pp 241--268, \mn@doi{10.1142/9789813227958_0013}

\bibitem[\protect\citeauthoryear{{Colpi} et~al.}{{Colpi}
  et~al.}{2019}]{Colpi19}
{Colpi} M.,  et~al., 2019, arXiv e-prints, \href
  {https://ui.adsabs.harvard.edu/abs/2019arXiv190306867C} {p. arXiv:1903.06867}

\bibitem[\protect\citeauthoryear{{Cowie}, {Barger}, {Bauer}  \&
  {Gonz{\'a}lez-L{\'o}pez}}{{Cowie} et~al.}{2020}]{Cowie20}
{Cowie} L.~L.,  {Barger} A.~J.,  {Bauer} F.~E.,   {Gonz{\'a}lez-L{\'o}pez} J.,
  2020, \mn@doi [\apj] {10.3847/1538-4357/ab6aaa}, \href
  {https://ui.adsabs.harvard.edu/abs/2020ApJ...891...69C} {891, 69}

\bibitem[\protect\citeauthoryear{{Dal Canton}, {Mangiagli}, {Noble},
  {Schnittman}, {Ptak}, {Klein}, {Sesana}  \& {Camp}}{{Dal Canton}
  et~al.}{2019}]{Dalcanton2019}
{Dal Canton} T.,  {Mangiagli} A.,  {Noble} S.~C.,  {Schnittman} J.,  {Ptak} A.,
   {Klein} A.,  {Sesana} A.,   {Camp} J.,  2019, \mn@doi [\apj]
  {10.3847/1538-4357/ab505a}, \href
  {https://ui.adsabs.harvard.edu/abs/2019ApJ...886..146D} {886, 146}

\bibitem[\protect\citeauthoryear{{Davies}, {Miller}  \& {Bellovary}}{{Davies}
  et~al.}{2011}]{Davies11}
{Davies} M.~B.,  {Miller} M.~C.,   {Bellovary} J.~M.,  2011, \mn@doi [\apj]
  {10.1088/2041-8205/740/2/L42}, \href
  {http://adsabs.harvard.edu/abs/2011ApJ...740L..42D} {740, L42}

\bibitem[\protect\citeauthoryear{{Dayal}, {Rossi}, {Shiralilou}, {Piana},
  {Choudhury}  \& {Volonteri}}{{Dayal} et~al.}{2019}]{Dayal2019}
{Dayal} P.,  {Rossi} E.~M.,  {Shiralilou} B.,  {Piana} O.,  {Choudhury} T.~R.,
   {Volonteri} M.,  2019, \mn@doi [\mnras] {10.1093/mnras/stz897}, \href
  {https://ui.adsabs.harvard.edu/abs/2019MNRAS.486.2336D} {486, 2336}

\bibitem[\protect\citeauthoryear{{Devecchi}, {Volonteri}, {Rossi}, {Colpi}  \&
  {Portegies Zwart}}{{Devecchi} et~al.}{2012}]{Devecchi12}
{Devecchi} B.,  {Volonteri} M.,  {Rossi} E.~M.,  {Colpi} M.,   {Portegies
  Zwart} S.,  2012, \mn@doi [\mnras] {10.1111/j.1365-2966.2012.20406.x}, \href
  {http://adsabs.harvard.edu/abs/2012MNRAS.421.1465D} {421, 1465}

\bibitem[\protect\citeauthoryear{{Di Carlo} et~al.,}{{Di Carlo}
  et~al.}{2020}]{DiCarlo2020}
{Di Carlo} U.~N.,  et~al., 2020, arXiv e-prints, \href
  {https://ui.adsabs.harvard.edu/abs/2020arXiv200409525D} {p. arXiv:2004.09525}

\bibitem[\protect\citeauthoryear{{Di Matteo}, {Springel}  \& {Hernquist}}{{Di
  Matteo} et~al.}{2005}]{DiMatteo05}
{Di Matteo} T.,  {Springel} V.,   {Hernquist} L.,  2005, \mn@doi [\nat]
  {10.1038/nature03335}, \href
  {http://adsabs.harvard.edu/abs/2005Natur.433..604D} {433, 604}

\bibitem[\protect\citeauthoryear{{Dijkstra}, {Ferrara}  \&
  {Mesinger}}{{Dijkstra} et~al.}{2014}]{Dijkstra14}
{Dijkstra} M.,  {Ferrara} A.,   {Mesinger} A.,  2014, \mn@doi [\mnras]
  {10.1093/mnras/stu1007}, \href
  {http://adsabs.harvard.edu/abs/2014MNRAS.442.2036D} {442, 2036}

\bibitem[\protect\citeauthoryear{{Dominik}, {Belczynski}, {Fryer}, {Holz},
  {Berti}, {Bulik}, {Mand el}  \& {O'Shaughnessy}}{{Dominik}
  et~al.}{2012}]{Dominick2012}
{Dominik} M.,  {Belczynski} K.,  {Fryer} C.,  {Holz} D.~E.,  {Berti} E.,
  {Bulik} T.,  {Mand el} I.,   {O'Shaughnessy} R.,  2012, \mn@doi [\apj]
  {10.1088/0004-637X/759/1/52}, \href
  {https://ui.adsabs.harvard.edu/abs/2012ApJ...759...52D} {759, 52}

\bibitem[\protect\citeauthoryear{{Dominik}, {Belczynski}, {Fryer}, {Holz},
  {Berti}, {Bulik}, {Mand el}  \& {O'Shaughnessy}}{{Dominik}
  et~al.}{2013}]{Dominick2013}
{Dominik} M.,  {Belczynski} K.,  {Fryer} C.,  {Holz} D.~E.,  {Berti} E.,
  {Bulik} T.,  {Mand el} I.,   {O'Shaughnessy} R.,  2013, \mn@doi [\apj]
  {10.1088/0004-637X/779/1/72}, \href
  {https://ui.adsabs.harvard.edu/abs/2013ApJ...779...72D} {779, 72}

\bibitem[\protect\citeauthoryear{{Dominik} et~al.,}{{Dominik}
  et~al.}{2015}]{Dominick2015}
{Dominik} M.,  et~al., 2015, \mn@doi [\apj] {10.1088/0004-637X/806/2/263},
  \href {https://ui.adsabs.harvard.edu/abs/2015ApJ...806..263D} {806, 263}

\bibitem[\protect\citeauthoryear{{Dosopoulou} \& {Antonini}}{{Dosopoulou} \&
  {Antonini}}{2017}]{Dosopoulou17}
{Dosopoulou} F.,  {Antonini} F.,  2017, \mn@doi [\apj]
  {10.3847/1538-4357/aa6b58}, \href
  {https://ui.adsabs.harvard.edu/abs/2017ApJ...840...31D} {840, 31}

\bibitem[\protect\citeauthoryear{{Enoki}, {Inoue}, {Nagashima}  \&
  {Sugiyama}}{{Enoki} et~al.}{2005}]{Enoki2005}
{Enoki} M.,  {Inoue} K.~T.,  {Nagashima} M.,   {Sugiyama} N.,  2005, Annual
  Report of the National Astronomical Observatory of Japan, \href
  {https://ui.adsabs.harvard.edu/abs/2005ARAOJ...7...34E} {7, 34}

\bibitem[\protect\citeauthoryear{{Fan} et~al.,}{{Fan} et~al.}{2001}]{Fan01}
{Fan} X.,  et~al., 2001, \mn@doi [\aj] {10.1086/324111}, \href
  {http://adsabs.harvard.edu/abs/2001AJ....122.2833F} {122, 2833}

\bibitem[\protect\citeauthoryear{{Fan}, {Hennawi}, {Richards}, {Strauss},
  {Schneider}, {Donley}, {Young}  \& {Annis}}{{Fan} et~al.}{2004}]{Fan04}
{Fan} X.,  {Hennawi} J.~F.,  {Richards} G.~T.,  {Strauss} M.~A.,  {Schneider}
  D.~P.,  {Donley} J.~L.,  {Young} J.~E.,   {Annis} e.~a.,  2004, \mn@doi [\aj]
  {10.1086/422434}, \href {http://adsabs.harvard.edu/abs/2004AJ....128..515F}
  {128, 515}

\bibitem[\protect\citeauthoryear{{Ferland} et~al.,}{{Ferland}
  et~al.}{2013}]{Ferland13}
{Ferland} G.~J.,  et~al., 2013, \rmxaa, \href
  {http://adsabs.harvard.edu/abs/2013RMxAA..49..137F} {49, 137}

\bibitem[\protect\citeauthoryear{{Fiacconi}, {Mayer}, {Ro{\v s}kar}  \&
  {Colpi}}{{Fiacconi} et~al.}{2013}]{Fiacconi13}
{Fiacconi} D.,  {Mayer} L.,  {Ro{\v s}kar} R.,   {Colpi} M.,  2013, \mn@doi
  [\apjl] {10.1088/2041-8205/777/1/L14}, \href
  {http://adsabs.harvard.edu/abs/2013ApJ...777L..14F} {777, L14}

\bibitem[\protect\citeauthoryear{{Graziani}, {Schneider}, {Marassi}, {Del
  Pozzo}, {Mapelli}  \& {Giacobbo}}{{Graziani} et~al.}{2020}]{Graziani2020}
{Graziani} L.,  {Schneider} R.,  {Marassi} S.,  {Del Pozzo} W.,  {Mapelli} M.,
   {Giacobbo} N.,  2020, \mn@doi [\mnras] {10.1093/mnrasl/slaa063}, \href
  {https://ui.adsabs.harvard.edu/abs/2020MNRAS.495L..81G} {495, L81}

\bibitem[\protect\citeauthoryear{{Habouzit}, {Volonteri}, {Latif}, {Dubois}  \&
  {Peirani}}{{Habouzit} et~al.}{2016}]{Habouzit16}
{Habouzit} M.,  {Volonteri} M.,  {Latif} M.,  {Dubois} Y.,   {Peirani} S.,
  2016, \mn@doi [\mnras] {10.1093/mnras/stw1924}, \href
  {http://adsabs.harvard.edu/abs/2016MNRAS.463..529H} {463, 529}

\bibitem[\protect\citeauthoryear{{Hall} \& {Evans}}{{Hall} \&
  {Evans}}{2019}]{HallEvans2019}
{Hall} E.~D.,  {Evans} M.,  2019, \mn@doi [Classical and Quantum Gravity]
  {10.1088/1361-6382/ab41d6}, \href
  {https://ui.adsabs.harvard.edu/abs/2019CQGra..36v5002H} {36, 225002}

\bibitem[\protect\citeauthoryear{{Heger} \& {Woosley}}{{Heger} \&
  {Woosley}}{2002}]{Heger02}
{Heger} A.,  {Woosley} S.~E.,  2002, \apj, 567, 532

\bibitem[\protect\citeauthoryear{{Heger} \& {Woosley}}{{Heger} \&
  {Woosley}}{2010}]{HegerWoosley10}
{Heger} A.,  {Woosley} S.~E.,  2010, \mn@doi [\apj]
  {10.1088/0004-637X/724/1/341}, \href
  {http://adsabs.harvard.edu/abs/2010ApJ...724..341H} {724, 341}

\bibitem[\protect\citeauthoryear{{Heger}, {Fryer}, {Woosley}, {Langer}  \&
  {Hartmann}}{{Heger} et~al.}{2003}]{Heger03}
{Heger} A.,  {Fryer} C.~L.,  {Woosley} S.~E.,  {Langer} N.,   {Hartmann} D.~H.,
   2003, \mn@doi [\apj] {10.1086/375341}, \href
  {http://adsabs.harvard.edu/abs/2003ApJ...591..288H} {591, 288}

\bibitem[\protect\citeauthoryear{{Hild} et~al.}{{Hild} et~al.}{2011}]{Hild2011}
{Hild} S.,  et~al., 2011, \mn@doi [Classical and Quantum Gravity]
  {10.1088/0264-9381/28/9/094013}, \href
  {https://ui.adsabs.harvard.edu/abs/2011CQGra..28i4013H} {28, 094013}

\bibitem[\protect\citeauthoryear{{Hirano}, {Hosokawa}, {Yoshida}, {Umeda},
  {Omukai}, {Chiaki}  \& {Yorke}}{{Hirano} et~al.}{2014}]{Hirano14}
{Hirano} S.,  {Hosokawa} T.,  {Yoshida} N.,  {Umeda} H.,  {Omukai} K.,
  {Chiaki} G.,   {Yorke} H.~W.,  2014, \mn@doi [\apj]
  {10.1088/0004-637X/781/2/60}, \href
  {http://adsabs.harvard.edu/abs/2014ApJ...781...60H} {781, 60}

\bibitem[\protect\citeauthoryear{{Hirano}, {Hosokawa}, {Yoshida}, {Omukai}  \&
  {Yorke}}{{Hirano} et~al.}{2015}]{Hirano15}
{Hirano} S.,  {Hosokawa} T.,  {Yoshida} N.,  {Omukai} K.,   {Yorke} H.~W.,
  2015, \mn@doi [\mnras] {10.1093/mnras/stv044}, \href
  {http://adsabs.harvard.edu/abs/2015MNRAS.448..568H} {448, 568}

\bibitem[\protect\citeauthoryear{{Hirano}, {Yoshida}, {Sakurai}  \&
  {Fujii}}{{Hirano} et~al.}{2018}]{Hirano2018}
{Hirano} S.,  {Yoshida} N.,  {Sakurai} Y.,   {Fujii} M.~S.,  2018, \mn@doi
  [\apj] {10.3847/1538-4357/aaaaba}, \href
  {http://adsabs.harvard.edu/abs/2018ApJ...855...17H} {855, 17}

\bibitem[\protect\citeauthoryear{{Huang}, {Feng}  \& {Di Matteo}}{{Huang}
  et~al.}{2019}]{Huang2019}
{Huang} K.-W.,  {Feng} Y.,   {Di Matteo} T.,  2019, arXiv e-prints, \href
  {https://ui.adsabs.harvard.edu/abs/2019arXiv190600242H} {p. arXiv:1906.00242}

\bibitem[\protect\citeauthoryear{{Inayoshi} \& {Omukai}}{{Inayoshi} \&
  {Omukai}}{2012}]{IO12}
{Inayoshi} K.,  {Omukai} K.,  2012, \mn@doi [\mnras]
  {10.1111/j.1365-2966.2012.20812.x}, \href
  {http://adsabs.harvard.edu/abs/2012MNRAS.422.2539I} {422, 2539}

\bibitem[\protect\citeauthoryear{{Inayoshi}, {Omukai}  \& {Tasker}}{{Inayoshi}
  et~al.}{2014}]{Inayoshi14}
{Inayoshi} K.,  {Omukai} K.,   {Tasker} E.,  2014, \mn@doi [\mnras]
  {10.1093/mnrasl/slu151}, \href
  {http://adsabs.harvard.edu/abs/2014MNRAS.445L.109I} {445, L109}

\bibitem[\protect\citeauthoryear{{Inayoshi}, {Hirai}, {Kinugawa}  \&
  {Hotokezaka}}{{Inayoshi} et~al.}{2017}]{Inayoshi17}
{Inayoshi} K.,  {Hirai} R.,  {Kinugawa} T.,   {Hotokezaka} K.,  2017, \mn@doi
  [\mnras] {10.1093/mnras/stx757}, \href
  {http://adsabs.harvard.edu/abs/2017MNRAS.468.5020I} {468, 5020}

\bibitem[\protect\citeauthoryear{{Inayoshi}, {Visbal}  \& {Haiman}}{{Inayoshi}
  et~al.}{2019}]{Inayoshi2019}
{Inayoshi} K.,  {Visbal} E.,   {Haiman} Z.,  2019, arXiv e-prints, \href
  {https://ui.adsabs.harvard.edu/abs/2019arXiv191105791I} {p. arXiv:1911.05791}

\bibitem[\protect\citeauthoryear{{Jani}, {Shoemaker}  \& {Cutler}}{{Jani}
  et~al.}{2019}]{Jani2019}
{Jani} K.,  {Shoemaker} D.,   {Cutler} C.,  2019, \mn@doi [Nature Astronomy]
  {10.1038/s41550-019-0932-7}, \href
  {https://ui.adsabs.harvard.edu/abs/2020NatAs...4..260J} {4, 260}

\bibitem[\protect\citeauthoryear{{Johnson} \& {Haardt}}{{Johnson} \&
  {Haardt}}{2016}]{Johnson16}
{Johnson} J.~L.,  {Haardt} F.,  2016, \mn@doi [\pasa] {10.1017/pasa.2016.4},
  \href {http://adsabs.harvard.edu/abs/2016PASA...33....7J} {33, e007}

\bibitem[\protect\citeauthoryear{{Kalogera} et~al.,}{{Kalogera}
  et~al.}{2019}]{Kalogera19}
{Kalogera} V.,  et~al., 2019, \baas, \href
  {https://ui.adsabs.harvard.edu/abs/2019BAAS...51c.242K} {51, 242}

\bibitem[\protect\citeauthoryear{{Katz}, {Kelley}, {Dosopoulou}, {Berry},
  {Blecha}  \& {Larson}}{{Katz} et~al.}{2020}]{Katz2020}
{Katz} M.~L.,  {Kelley} L.~Z.,  {Dosopoulou} F.,  {Berry} S.,  {Blecha} L.,
  {Larson} S.~L.,  2020, \mn@doi [\mnras] {10.1093/mnras/stz3102}, \href
  {https://ui.adsabs.harvard.edu/abs/2020MNRAS.491.2301K} {491, 2301}

\bibitem[\protect\citeauthoryear{{Kelley}, {Blecha}  \& {Hernquist}}{{Kelley}
  et~al.}{2017}]{2017MNRAS.464.3131K}
{Kelley} L.~Z.,  {Blecha} L.,   {Hernquist} L.,  2017, \mn@doi [\mnras]
  {10.1093/mnras/stw2452}, \href
  {https://ui.adsabs.harvard.edu/abs/2017MNRAS.464.3131K} {464, 3131}

\bibitem[\protect\citeauthoryear{{Khan}, {Fiacconi}, {Mayer}, {Berczik}  \&
  {Just}}{{Khan} et~al.}{2016}]{Khan16}
{Khan} F.~M.,  {Fiacconi} D.,  {Mayer} L.,  {Berczik} P.,   {Just} A.,  2016,
  \mn@doi [\apj] {10.3847/0004-637X/828/2/73}, \href
  {https://ui.adsabs.harvard.edu/abs/2016ApJ...828...73K} {828, 73}

\bibitem[\protect\citeauthoryear{{Klein} et~al.,}{{Klein}
  et~al.}{2016}]{Klein2016}
{Klein} A.,  et~al., 2016, \mn@doi [\prd] {10.1103/PhysRevD.93.024003}, \href
  {https://ui.adsabs.harvard.edu/abs/2016PhRvD..93b4003K} {93, 024003}

\bibitem[\protect\citeauthoryear{{Kormendy} \& {Ho}}{{Kormendy} \&
  {Ho}}{2013}]{Kormendy13}
{Kormendy} J.,  {Ho} L.~C.,  2013, \mn@doi [\araa]
  {10.1146/annurev-astro-082708-101811}, \href
  {http://adsabs.harvard.edu/abs/2013ARA\%26A..51..511K} {51, 511}

\bibitem[\protect\citeauthoryear{{Kozai}}{{Kozai}}{1962}]{kozai1962}
{Kozai} Y.,  1962, \mn@doi [\aj] {10.1086/108790}, \href
  {https://ui.adsabs.harvard.edu/abs/1962AJ.....67..591K} {67, 591}

\bibitem[\protect\citeauthoryear{{Larson}}{{Larson}}{1998}]{Larson98}
{Larson} R.~B.,  1998, \mnras, 301, 569

\bibitem[\protect\citeauthoryear{{Latif} \& {Ferrara}}{{Latif} \&
  {Ferrara}}{2016}]{Latif16}
{Latif} M.~A.,  {Ferrara} A.,  2016, \mn@doi [\pasa] {10.1017/pasa.2016.41},
  \href {http://adsabs.harvard.edu/abs/2016PASA...33...51L} {33, e051}

\bibitem[\protect\citeauthoryear{{Latif}, {Schleicher}, {Schmidt}  \&
  {Niemeyer}}{{Latif} et~al.}{2013}]{Latif13}
{Latif} M.~A.,  {Schleicher} D.~R.~G.,  {Schmidt} W.,   {Niemeyer} J.~C.,
  2013, \mn@doi [\mnras] {10.1093/mnras/stt1786}, \href
  {http://adsabs.harvard.edu/abs/2013MNRAS.436.2989L} {436, 2989}

\bibitem[\protect\citeauthoryear{{Lidov}}{{Lidov}}{1962}]{Lidov1962}
{Lidov} M.~L.,  1962, \mn@doi [\planss] {10.1016/0032-0633(62)90129-0}, \href
  {https://ui.adsabs.harvard.edu/abs/1962P&SS....9..719L} {9, 719}

\bibitem[\protect\citeauthoryear{{Luo} et~al.,}{{Luo} et~al.}{2016}]{Luo2016}
{Luo} J.,  et~al., 2016, \mn@doi [Classical and Quantum Gravity]
  {10.1088/0264-9381/33/3/035010}, \href
  {https://ui.adsabs.harvard.edu/abs/2016CQGra..33c5010L} {33, 035010}

\bibitem[\protect\citeauthoryear{{Lupi}, {Colpi}, {Devecchi}, {Galanti}  \&
  {Volonteri}}{{Lupi} et~al.}{2014}]{Lupi14}
{Lupi} A.,  {Colpi} M.,  {Devecchi} B.,  {Galanti} G.,   {Volonteri} M.,  2014,
  \mn@doi [\mnras] {10.1093/mnras/stu1120}, \href
  {https://ui.adsabs.harvard.edu/abs/2014MNRAS.442.3616L} {442, 3616}

\bibitem[\protect\citeauthoryear{{Madau} \& {Rees}}{{Madau} \&
  {Rees}}{2001}]{MadauRees01}
{Madau} P.,  {Rees} M.~J.,  2001, \mn@doi [\apjl] {10.1086/319848}, \href
  {http://adsabs.harvard.edu/abs/2001ApJ...551L..27M} {551, L27}

\bibitem[\protect\citeauthoryear{{Madau}, {Haardt}  \& {Dotti}}{{Madau}
  et~al.}{2014}]{2014ApJ...784L..38M}
{Madau} P.,  {Haardt} F.,   {Dotti} M.,  2014, \mn@doi [\apjl]
  {10.1088/2041-8205/784/2/L38}, \href
  {https://ui.adsabs.harvard.edu/abs/2014ApJ...784L..38M} {784, L38}

\bibitem[\protect\citeauthoryear{{Maggiore} et~al.,}{{Maggiore}
  et~al.}{2019}]{Maggiore19}
{Maggiore} M.,  et~al., 2019, arXiv e-prints, \href
  {https://ui.adsabs.harvard.edu/abs/2019arXiv191202622M} {p. arXiv:1912.02622}

\bibitem[\protect\citeauthoryear{{Mapelli}}{{Mapelli}}{2016}]{Mapelli16}
{Mapelli} M.,  2016, \mn@doi [\mnras] {10.1093/mnras/stw869}, \href
  {http://adsabs.harvard.edu/abs/2016MNRAS.459.3432M} {459, 3432}

\bibitem[\protect\citeauthoryear{{Mapelli} \& {Giacobbo}}{{Mapelli} \&
  {Giacobbo}}{2018}]{Mapelli2018}
{Mapelli} M.,  {Giacobbo} N.,  2018, \mn@doi [\mnras] {10.1093/mnras/sty1613},
  \href {https://ui.adsabs.harvard.edu/abs/2018MNRAS.479.4391M} {479, 4391}

\bibitem[\protect\citeauthoryear{{Mapelli}, {Giacobbo}, {Ripamonti}  \&
  {Spera}}{{Mapelli} et~al.}{2017}]{Mapelli2017}
{Mapelli} M.,  {Giacobbo} N.,  {Ripamonti} E.,   {Spera} M.,  2017, \mn@doi
  [\mnras] {10.1093/mnras/stx2123}, \href
  {https://ui.adsabs.harvard.edu/abs/2017MNRAS.472.2422M} {472, 2422}

\bibitem[\protect\citeauthoryear{{Mapelli}, {Giacobbo}, {Santoliquido}  \&
  {Artale}}{{Mapelli} et~al.}{2019}]{Mapelli2019}
{Mapelli} M.,  {Giacobbo} N.,  {Santoliquido} F.,   {Artale} M.~C.,  2019,
  \mn@doi [\mnras] {10.1093/mnras/stz1150}, \href
  {https://ui.adsabs.harvard.edu/abs/2019MNRAS.487....2M} {487, 2}

\bibitem[\protect\citeauthoryear{{Marassi}, {Graziani}, {Ginolfi}, {Schneider},
  {Mapelli}, {Spera}  \& {Alparone}}{{Marassi} et~al.}{2019}]{Marassi2019}
{Marassi} S.,  {Graziani} L.,  {Ginolfi} M.,  {Schneider} R.,  {Mapelli} M.,
  {Spera} M.,   {Alparone} M.,  2019, \mn@doi [\mnras] {10.1093/mnras/stz170},
  \href {https://ui.adsabs.harvard.edu/abs/2019MNRAS.484.3219M} {484, 3219}

\bibitem[\protect\citeauthoryear{{Marconi}, {Risaliti}, {Gilli}, {Hunt},
  {Maiolino}  \& {Salvati}}{{Marconi} et~al.}{2004}]{Marconi04}
{Marconi} A.,  {Risaliti} G.,  {Gilli} R.,  {Hunt} L.~K.,  {Maiolino} R.,
  {Salvati} M.,  2004, \mn@doi [\mnras] {10.1111/j.1365-2966.2004.07765.x},
  \href {http://adsabs.harvard.edu/abs/2004MNRAS.351..169M} {351, 169}

\bibitem[\protect\citeauthoryear{{Matsuoka} et~al.,}{{Matsuoka}
  et~al.}{2018}]{Matsuoka18}
{Matsuoka} Y.,  et~al., 2018, \mn@doi [\apj] {10.3847/1538-4357/aaee7a}, \href
  {http://adsabs.harvard.edu/abs/2018ApJ...869..150M} {869, 150}

\bibitem[\protect\citeauthoryear{{Mayer}, {Kazantzidis}, {Madau}, {Colpi},
  {Quinn}  \& {Wadsley}}{{Mayer} et~al.}{2007}]{Mayer07}
{Mayer} L.,  {Kazantzidis} S.,  {Madau} P.,  {Colpi} M.,  {Quinn} T.,
  {Wadsley} J.,  2007, \mn@doi [Science] {10.1126/science.1141858}, \href
  {http://adsabs.harvard.edu/abs/2007Sci...316.1874M} {316, 1874}

\bibitem[\protect\citeauthoryear{{Mayer}, {Fiacconi}, {Bonoli}, {Quinn}, {Ro{\v
  s}kar}, {Shen}  \& {Wadsley}}{{Mayer} et~al.}{2015}]{Mayer15}
{Mayer} L.,  {Fiacconi} D.,  {Bonoli} S.,  {Quinn} T.,  {Ro{\v s}kar} R.,
  {Shen} S.,   {Wadsley} J.,  2015, \mn@doi [\apj]
  {10.1088/0004-637X/810/1/51}, \href
  {http://adsabs.harvard.edu/abs/2015ApJ...810...51M} {810, 51}

\bibitem[\protect\citeauthoryear{{McGee}, {Sesana}  \& {Vecchio}}{{McGee}
  et~al.}{2020}]{SesanaNature2020}
{McGee} S.,  {Sesana} A.,   {Vecchio} A.,  2020, \mn@doi [Nature Astronomy]
  {10.1038/s41550-019-0969-7}, \href
  {https://ui.adsabs.harvard.edu/abs/2020NatAs...4...26M} {4, 26}

\bibitem[\protect\citeauthoryear{{Merloni}, {Rudnick}  \& {Di
  Matteo}}{{Merloni} et~al.}{2004}]{Merloni04b}
{Merloni} A.,  {Rudnick} G.,   {Di Matteo} T.,  2004, \mn@doi [\mnras]
  {10.1111/j.1365-2966.2004.08382.x}, \href
  {http://adsabs.harvard.edu/abs/2004MNRAS.354L..37M} {354, L37}

\bibitem[\protect\citeauthoryear{{Mezcua}, {Civano}, {Fabbiano}, {Miyaji}  \&
  {Marchesi}}{{Mezcua} et~al.}{2016}]{Mezcua2016}
{Mezcua} M.,  {Civano} F.,  {Fabbiano} G.,  {Miyaji} T.,   {Marchesi} S.,
  2016, \mn@doi [\apj] {10.3847/0004-637X/817/1/20}, \href
  {https://ui.adsabs.harvard.edu/abs/2016ApJ...817...20M} {817, 20}

\bibitem[\protect\citeauthoryear{{Mezcua}, {Civano}, {Marchesi}, {Suh},
  {Fabbiano}  \& {Volonteri}}{{Mezcua} et~al.}{2018}]{Mezcua2018}
{Mezcua} M.,  {Civano} F.,  {Marchesi} S.,  {Suh} H.,  {Fabbiano} G.,
  {Volonteri} M.,  2018, \mn@doi [\mnras] {10.1093/mnras/sty1163}, \href
  {https://ui.adsabs.harvard.edu/abs/2018MNRAS.478.2576M} {478, 2576}

\bibitem[\protect\citeauthoryear{{Nardini} et~al.,}{{Nardini}
  et~al.}{2015}]{Nardini15}
{Nardini} E.,  et~al., 2015, \mn@doi [Science] {10.1126/science.1259202}, \href
  {https://ui.adsabs.harvard.edu/abs/2015Sci...347..860N} {347, 860}

\bibitem[\protect\citeauthoryear{{Natarajan}, {Pacucci}, {Ferrara}, {Agarwal},
  {Ricarte}, {Zackrisson}  \& {Cappelluti}}{{Natarajan}
  et~al.}{2017}]{Natarajan17}
{Natarajan} P.,  {Pacucci} F.,  {Ferrara} A.,  {Agarwal} B.,  {Ricarte} A.,
  {Zackrisson} E.,   {Cappelluti} N.,  2017, \mn@doi [\apj]
  {10.3847/1538-4357/aa6330}, \href
  {http://adsabs.harvard.edu/abs/2017ApJ...838..117N} {838, 117}

\bibitem[\protect\citeauthoryear{{Pacucci}, {Ferrara}, {Volonteri}  \&
  {Dubus}}{{Pacucci} et~al.}{2015}]{Pacucci15}
{Pacucci} F.,  {Ferrara} A.,  {Volonteri} M.,   {Dubus} G.,  2015, \mn@doi
  [\mnras] {10.1093/mnras/stv2196}, \href
  {http://adsabs.harvard.edu/abs/2015MNRAS.454.3771P} {454, 3771}

\bibitem[\protect\citeauthoryear{{Pezzulli}, {Volonteri}, {Schneider}  \&
  {Valiante}}{{Pezzulli} et~al.}{2017a}]{Pezzulli17}
{Pezzulli} E.,  {Volonteri} M.,  {Schneider} R.,   {Valiante} R.,  2017a,
  \mn@doi [\mnras] {10.1093/mnras/stx1640}, \href
  {http://adsabs.harvard.edu/abs/2017MNRAS.471..589P} {471, 589}

\bibitem[\protect\citeauthoryear{{Pezzulli}, {Volonteri}, {Schneider}  \&
  {Valiante}}{{Pezzulli} et~al.}{2017b}]{Pezzulli16}
{Pezzulli} E.,  {Volonteri} M.,  {Schneider} R.,   {Valiante} R.,  2017b,
  \mn@doi [\mnras] {10.1093/mnras/stx1640}, \href
  {http://adsabs.harvard.edu/abs/2017MNRAS.471..589P} {471, 589}

\bibitem[\protect\citeauthoryear{{Pfister}, {Lupi}, {Capelo}, {Volonteri},
  {Bellovary}  \& {Dotti}}{{Pfister} et~al.}{2017}]{Pfister2017}
{Pfister} H.,  {Lupi} A.,  {Capelo} P.~R.,  {Volonteri} M.,  {Bellovary} J.~M.,
    {Dotti} M.,  2017, \mn@doi [\mnras] {10.1093/mnras/stx1853}, \href
  {https://ui.adsabs.harvard.edu/abs/2017MNRAS.471.3646P} {471, 3646}

\bibitem[\protect\citeauthoryear{{Pfister}, {Volonteri}, {Dubois}, {Dotti}  \&
  {Colpi}}{{Pfister} et~al.}{2019}]{Pfister2019}
{Pfister} H.,  {Volonteri} M.,  {Dubois} Y.,  {Dotti} M.,   {Colpi} M.,  2019,
  \mn@doi [\mnras] {10.1093/mnras/stz822}, \href
  {https://ui.adsabs.harvard.edu/abs/2019MNRAS.486..101P} {486, 101}

\bibitem[\protect\citeauthoryear{{Press} \& {Schechter}}{{Press} \&
  {Schechter}}{1974}]{PS74}
{Press} W.~H.,  {Schechter} P.,  1974, \mn@doi [\apj] {10.1086/152650}, \href
  {http://adsabs.harvard.edu/abs/1974ApJ...187..425P} {187, 425}

\bibitem[\protect\citeauthoryear{{Punturo} et~al.}{{Punturo}
  et~al.}{2010}]{ET10}
{Punturo} M.,  et~al., 2010, \mn@doi [Classical and Quantum Gravity]
  {10.1088/0264-9381/27/19/194002}, \href
  {https://ui.adsabs.harvard.edu/abs/2010CQGra..27s4002P} {27, 194002}

\bibitem[\protect\citeauthoryear{{Regan}, {Visbal}, {Wise}, {Haiman},
  {Johansson}  \& {Bryan}}{{Regan} et~al.}{2017}]{Regan17}
{Regan} J.~A.,  {Visbal} E.,  {Wise} J.~H.,  {Haiman} Z.,  {Johansson} P.~H.,
  {Bryan} G.~L.,  2017, \mn@doi [Nature Astronomy] {10.1038/s41550-017-0075},
  \href {http://adsabs.harvard.edu/abs/2017NatAs...1E..75R} {1, 0075}

\bibitem[\protect\citeauthoryear{{Reines} \& {Volonteri}}{{Reines} \&
  {Volonteri}}{2015}]{Reines15}
{Reines} A.~E.,  {Volonteri} M.,  2015, \mn@doi [\apj]
  {10.1088/0004-637X/813/2/82}, \href
  {http://adsabs.harvard.edu/abs/2015ApJ...813...82R} {813, 82}

\bibitem[\protect\citeauthoryear{{Reinoso}, {Schleicher}, {Fellhauer},
  {Klessen}  \& {Boekholt}}{{Reinoso} et~al.}{2018}]{Reinoso18}
{Reinoso} B.,  {Schleicher} D.~R.~G.,  {Fellhauer} M.,  {Klessen} R.~S.,
  {Boekholt} T.~C.~N.,  2018, \mn@doi [\aap] {10.1051/0004-6361/201732224},
  \href {http://adsabs.harvard.edu/abs/2018A%26A...614A..14R} {614, A14}

\bibitem[\protect\citeauthoryear{{Reitze} et~al.,}{{Reitze}
  et~al.}{2019}]{Reitze2019}
{Reitze} D.,  et~al., 2019, in \baas. p.~35 (\mn@eprint {arXiv} {1907.04833})

\bibitem[\protect\citeauthoryear{{Ricarte} \& {Natarajan}}{{Ricarte} \&
  {Natarajan}}{2018}]{Ricarte2018}
{Ricarte} A.,  {Natarajan} P.,  2018, \mn@doi [\mnras] {10.1093/mnras/sty2448},
  \href {https://ui.adsabs.harvard.edu/abs/2018MNRAS.481.3278R} {481, 3278}

\bibitem[\protect\citeauthoryear{Robson, Cornish  \& Liu}{Robson
  et~al.}{2019}]{Robson_2019}
Robson T.,  Cornish N.~J.,   Liu C.,  2019, \mn@doi [Classical and Quantum
  Gravity] {10.1088/1361-6382/ab1101}, 36, 105011

\bibitem[\protect\citeauthoryear{{Rodriguez}, {Chatterjee}  \&
  {Rasio}}{{Rodriguez} et~al.}{2016}]{Rodriguez2016}
{Rodriguez} C.~L.,  {Chatterjee} S.,   {Rasio} F.~A.,  2016, \mn@doi [\prd]
  {10.1103/PhysRevD.93.084029}, \href
  {https://ui.adsabs.harvard.edu/abs/2016PhRvD..93h4029R} {93, 084029}

\bibitem[\protect\citeauthoryear{{Ruan}, {Guo}, {Cai}  \& {Zhang}}{{Ruan}
  et~al.}{2018}]{Ruan2018}
{Ruan} W.-H.,  {Guo} Z.-K.,  {Cai} R.-G.,   {Zhang} Y.-Z.,  2018, arXiv
  e-prints, \href {https://ui.adsabs.harvard.edu/abs/2018arXiv180709495R} {p.
  arXiv:1807.09495}

\bibitem[\protect\citeauthoryear{{Santamar{\'{\i}}a}
  et~al.,}{{Santamar{\'{\i}}a} et~al.}{2010}]{Santamaria10}
{Santamar{\'{\i}}a} L.,  et~al., 2010, \mn@doi [\prd]
  {10.1103/PhysRevD.82.064016}, \href
  {http://adsabs.harvard.edu/abs/2010PhRvD..82f4016S} {82, 064016}

\bibitem[\protect\citeauthoryear{{Santoliquido}, {Mapelli}, {Bouffanais},
  {Giacobbo}, {Di Carlo}, {Rastello}, {Artale}  \& {Ballone}}{{Santoliquido}
  et~al.}{2020}]{Santoliquido2020}
{Santoliquido} F.,  {Mapelli} M.,  {Bouffanais} Y.,  {Giacobbo} N.,  {Di Carlo}
  U.~N.,  {Rastello} S.,  {Artale} M.~C.,   {Ballone} A.,  2020, arXiv
  e-prints, \href {https://ui.adsabs.harvard.edu/abs/2020arXiv200409533S} {p.
  arXiv:2004.09533}

\bibitem[\protect\citeauthoryear{{Sathyaprakash} et~al.,}{{Sathyaprakash}
  et~al.}{2012}]{ET12}
{Sathyaprakash} B.,  et~al., 2012, \mn@doi [Classical and Quantum Gravity]
  {10.1088/0264-9381/29/12/124013}, \href
  {http://adsabs.harvard.edu/abs/2012CQGra..29l4013S} {29, 124013}

\bibitem[\protect\citeauthoryear{{Sato} et~al.,}{{Sato}
  et~al.}{2017}]{Sato2017}
{Sato} S.,  et~al., 2017, in Journal of Physics Conference Series. p. 012010,
  \mn@doi{10.1088/1742-6596/840/1/012010}

\bibitem[\protect\citeauthoryear{{Schleicher}, {Palla}, {Ferrara}, {Galli}  \&
  {Latif}}{{Schleicher} et~al.}{2013}]{Schleicher13}
{Schleicher} D.~R.~G.,  {Palla} F.,  {Ferrara} A.,  {Galli} D.,   {Latif} M.,
  2013, \mn@doi [\aap] {10.1051/0004-6361/201321949}, \href
  {http://adsabs.harvard.edu/abs/2013A%26A...558A..59S} {558, A59}

\bibitem[\protect\citeauthoryear{{Schneider}, {Ferrara}, {Natarajan}  \&
  {Omukai}}{{Schneider} et~al.}{2002}]{Schneider02}
{Schneider} R.,  {Ferrara} A.,  {Natarajan} P.,   {Omukai} K.,  2002, \mn@doi
  [\apj] {10.1086/339917}, \href
  {http://adsabs.harvard.edu/abs/2002ApJ...571...30S} {571, 30}

\bibitem[\protect\citeauthoryear{{Schneider}, {Ferrara}, {Salvaterra}, {Omukai}
   \& {Bromm}}{{Schneider} et~al.}{2003}]{Schneider03}
{Schneider} R.,  {Ferrara} A.,  {Salvaterra} R.,  {Omukai} K.,   {Bromm} V.,
  2003, \nat, \href {http://adsabs.harvard.edu/abs/2003Natur.422..869S} {422,
  869}

\bibitem[\protect\citeauthoryear{{Schneider}, {Omukai}, {Bianchi}  \&
  {Valiante}}{{Schneider} et~al.}{2012}]{Schneider12}
{Schneider} R.,  {Omukai} K.,  {Bianchi} S.,   {Valiante} R.,  2012, \mn@doi
  [\mnras] {10.1111/j.1365-2966.2011.19818.x}, \href
  {http://adsabs.harvard.edu/abs/2012MNRAS.419.1566S} {419, 1566}

\bibitem[\protect\citeauthoryear{{Schneider}, {Graziani}, {Marassi}, {Spera},
  {Mapelli}, {Alparone}  \& {Bennassuti}}{{Schneider}
  et~al.}{2017}]{Schneider2017}
{Schneider} R.,  {Graziani} L.,  {Marassi} S.,  {Spera} M.,  {Mapelli} M.,
  {Alparone} M.,   {Bennassuti} M.~d.,  2017, \mn@doi [\mnras]
  {10.1093/mnrasl/slx118}, \href
  {https://ui.adsabs.harvard.edu/abs/2017MNRAS.471L.105S} {471, L105}

\bibitem[\protect\citeauthoryear{{Schulze} et~al.,}{{Schulze}
  et~al.}{2019}]{Schulze19}
{Schulze} A.,  et~al., 2019, \mn@doi [\mnras] {10.1093/mnras/stz1746}, \href
  {https://ui.adsabs.harvard.edu/abs/2019MNRAS.488.1180S} {488, 1180}

\bibitem[\protect\citeauthoryear{{Sesana}, {Volonteri}  \& {Haardt}}{{Sesana}
  et~al.}{2007a}]{Sesana07}
{Sesana} A.,  {Volonteri} M.,   {Haardt} F.,  2007a, \mn@doi [\mnras]
  {10.1111/j.1365-2966.2007.11734.x}, \href
  {http://adsabs.harvard.edu/abs/2007MNRAS.377.1711S} {377, 1711}

\bibitem[\protect\citeauthoryear{{Sesana}, {Volonteri}  \& {Haardt}}{{Sesana}
  et~al.}{2007b}]{Sesana2007}
{Sesana} A.,  {Volonteri} M.,   {Haardt} F.,  2007b, \mn@doi [\mnras]
  {10.1111/j.1365-2966.2007.11734.x}, \href
  {https://ui.adsabs.harvard.edu/abs/2007MNRAS.377.1711S} {377, 1711}

\bibitem[\protect\citeauthoryear{{Sesana}, {Gair}, {Berti}  \&
  {Volonteri}}{{Sesana} et~al.}{2011}]{SesanaGair11}
{Sesana} A.,  {Gair} J.,  {Berti} E.,   {Volonteri} M.,  2011, \mn@doi [\prd]
  {10.1103/PhysRevD.83.044036}, \href
  {http://adsabs.harvard.edu/abs/2011PhRvD..83d4036S} {83, 044036}

\bibitem[\protect\citeauthoryear{{Shen} et~al.,}{{Shen} et~al.}{2011}]{Shen11}
{Shen} Y.,  et~al., 2011, \mn@doi [\apjs] {10.1088/0067-0049/194/2/45}, \href
  {https://ui.adsabs.harvard.edu/abs/2011ApJS..194...45S} {194, 45}

\bibitem[\protect\citeauthoryear{{Souza Lima}, {Mayer}, {Capelo}, {Bortolas}
  \& {Quinn}}{{Souza Lima} et~al.}{2020}]{Lima2020}
{Souza Lima} R.,  {Mayer} L.,  {Capelo} P.~R.,  {Bortolas} E.,   {Quinn} T.~R.,
   2020, arXiv e-prints, \href
  {https://ui.adsabs.harvard.edu/abs/2020arXiv200313789S} {p. arXiv:2003.13789}

\bibitem[\protect\citeauthoryear{{Sugimura}, {Matsumoto}, {Hosokawa}, {Hirano}
  \& {Omukai}}{{Sugimura} et~al.}{2020}]{Sugimura2020}
{Sugimura} K.,  {Matsumoto} T.,  {Hosokawa} T.,  {Hirano} S.,   {Omukai} K.,
  2020, \mn@doi [\apjl] {10.3847/2041-8213/ab7d37}, \href
  {https://ui.adsabs.harvard.edu/abs/2020ApJ...892L..14S} {892, L14}

\bibitem[\protect\citeauthoryear{{Tamanini}, {Caprini}, {Barausse}, {Sesana},
  {Klein}  \& {Petiteau}}{{Tamanini} et~al.}{2016}]{Tamanini16}
{Tamanini} N.,  {Caprini} C.,  {Barausse} E.,  {Sesana} A.,  {Klein} A.,
  {Petiteau} A.,  2016, \mn@doi [\jcap] {10.1088/1475-7516/2016/04/002}, \href
  {http://adsabs.harvard.edu/abs/2016JCAP...04..002T} {4, 002}

\bibitem[\protect\citeauthoryear{{Tamburello}, {Capelo}, {Mayer}, {Bellovary}
  \& {Wadsley}}{{Tamburello} et~al.}{2017}]{Tamburello2017}
{Tamburello} V.,  {Capelo} P.~R.,  {Mayer} L.,  {Bellovary} J.~M.,   {Wadsley}
  J.~W.,  2017, \mn@doi [\mnras] {10.1093/mnras/stw2561}, \href
  {https://ui.adsabs.harvard.edu/abs/2017MNRAS.464.2952T} {464, 2952}

\bibitem[\protect\citeauthoryear{{Tamfal}, {Capelo}, {Kazantzidis}, {Mayer},
  {Potter}, {Stadel}  \& {Widrow}}{{Tamfal} et~al.}{2018}]{Tamfal2018}
{Tamfal} T.,  {Capelo} P.~R.,  {Kazantzidis} S.,  {Mayer} L.,  {Potter} D.,
  {Stadel} J.,   {Widrow} L.~M.,  2018, \mn@doi [\apjl]
  {10.3847/2041-8213/aada4b}, \href
  {https://ui.adsabs.harvard.edu/abs/2018ApJ...864L..19T} {864, L19}

\bibitem[\protect\citeauthoryear{{Trakhtenbrot}}{{Trakhtenbrot}}{2020}]{Trakhtenbrot2020review}
{Trakhtenbrot} B.,  2020, arXiv e-prints, \href
  {https://ui.adsabs.harvard.edu/abs/2020arXiv200200972T} {p. arXiv:2002.00972}

\bibitem[\protect\citeauthoryear{{Umeda}, {Hosokawa}, {Omukai}  \&
  {Yoshida}}{{Umeda} et~al.}{2016}]{Umeda16}
{Umeda} H.,  {Hosokawa} T.,  {Omukai} K.,   {Yoshida} N.,  2016, \mn@doi
  [\apjl] {10.3847/2041-8205/830/2/L34}, \href
  {http://adsabs.harvard.edu/abs/2016ApJ...830L..34U} {830, L34}

\bibitem[\protect\citeauthoryear{{Valiante}, {Schneider}, {Salvadori}  \&
  {Bianchi}}{{Valiante} et~al.}{2011}]{Valiante11}
{Valiante} R.,  {Schneider} R.,  {Salvadori} S.,   {Bianchi} S.,  2011, \mn@doi
  [\mnras] {10.1111/j.1365-2966.2011.19168.x}, \href
  {http://adsabs.harvard.edu/abs/2011MNRAS.416.1916V} {416, 1916}

\bibitem[\protect\citeauthoryear{{Valiante}, {Schneider}, {Salvadori}  \&
  {Gallerani}}{{Valiante} et~al.}{2014}]{Valiante14}
{Valiante} R.,  {Schneider} R.,  {Salvadori} S.,   {Gallerani} S.,  2014,
  \mn@doi [\mnras] {10.1093/mnras/stu1613}, \href
  {http://adsabs.harvard.edu/abs/2014MNRAS.444.2442V} {444, 2442}

\bibitem[\protect\citeauthoryear{{Valiante}, {Schneider}, {Volonteri}  \&
  {Omukai}}{{Valiante} et~al.}{2016}]{Valiante16}
{Valiante} R.,  {Schneider} R.,  {Volonteri} M.,   {Omukai} K.,  2016, \mn@doi
  [\mnras] {10.1093/mnras/stw225}, \href
  {http://adsabs.harvard.edu/abs/2016MNRAS.457.3356V} {457, 3356}

\bibitem[\protect\citeauthoryear{{Valiante}, {Agarwal}, {Habouzit}  \&
  {Pezzulli}}{{Valiante} et~al.}{2017}]{Valiante17review}
{Valiante} R.,  {Agarwal} B.,  {Habouzit} M.,   {Pezzulli} E.,  2017, \mn@doi
  [\pasa] {10.1017/pasa.2017.25}, \href
  {http://adsabs.harvard.edu/abs/2017PASA...34...31V} {34, e031}

\bibitem[\protect\citeauthoryear{{Valiante}, {Schneider}, {Graziani}  \&
  {Zappacosta}}{{Valiante} et~al.}{2018a}]{Valiante18a}
{Valiante} R.,  {Schneider} R.,  {Graziani} L.,   {Zappacosta} L.,  2018a,
  \mn@doi [\mnras] {10.1093/mnras/stx3028}, \href
  {http://adsabs.harvard.edu/abs/2018MNRAS.474.3825V} {474, 3825}

\bibitem[\protect\citeauthoryear{{Valiante}, {Schneider}, {Zappacosta},
  {Graziani}, {Pezzulli}  \& {Volonteri}}{{Valiante}
  et~al.}{2018b}]{Valiante18b}
{Valiante} R.,  {Schneider} R.,  {Zappacosta} L.,  {Graziani} L.,  {Pezzulli}
  E.,   {Volonteri} M.,  2018b, \mn@doi [\mnras] {10.1093/mnras/sty213}, \href
  {http://adsabs.harvard.edu/abs/2018MNRAS.476..407V} {476, 407}

\bibitem[\protect\citeauthoryear{{Van Wassenhove}, {Capelo}, {Volonteri},
  {Dotti}, {Bellovary}, {Mayer}  \& {Governato}}{{Van Wassenhove}
  et~al.}{2014}]{Van2014}
{Van Wassenhove} S.,  {Capelo} P.~R.,  {Volonteri} M.,  {Dotti} M.,
  {Bellovary} J.~M.,  {Mayer} L.,   {Governato} F.,  2014, \mn@doi [\mnras]
  {10.1093/mnras/stu024}, \href
  {https://ui.adsabs.harvard.edu/abs/2014MNRAS.439..474V} {439, 474}

\bibitem[\protect\citeauthoryear{{Volonteri}}{{Volonteri}}{2010}]{Volonteri10}
{Volonteri} M.,  2010, \mn@doi [\aapr] {10.1007/s00159-010-0029-x}, \href
  {http://adsabs.harvard.edu/abs/2010A%26ARv..18..279V} {18, 279}

\bibitem[\protect\citeauthoryear{{Volonteri}, {Haardt}  \& {Madau}}{{Volonteri}
  et~al.}{2003}]{Volonteri03}
{Volonteri} M.,  {Haardt} F.,   {Madau} P.,  2003, \mn@doi [\apj]
  {10.1086/344675}, \href {http://adsabs.harvard.edu/abs/2003ApJ...582..559V}
  {582, 559}

\bibitem[\protect\citeauthoryear{{Volonteri}, {Silk}  \& {Dubus}}{{Volonteri}
  et~al.}{2015}]{2015ApJ...804..148V}
{Volonteri} M.,  {Silk} J.,   {Dubus} G.,  2015, \mn@doi [\apj]
  {10.1088/0004-637X/804/2/148}, \href
  {https://ui.adsabs.harvard.edu/abs/2015ApJ...804..148V} {804, 148}

\bibitem[\protect\citeauthoryear{{Volonteri}, {Reines}, {Atek}, {Stark}  \&
  {Trebitsch}}{{Volonteri} et~al.}{2017}]{Volonteri17}
{Volonteri} M.,  {Reines} A.,  {Atek} H.,  {Stark} D.~P.,   {Trebitsch} M.,
  2017, preprint, \href {http://adsabs.harvard.edu/abs/2017arXiv170400753V} {}
  (\mn@eprint {arXiv} {1704.00753})

\bibitem[\protect\citeauthoryear{{Volonteri} et~al.,}{{Volonteri}
  et~al.}{2020}]{Volonteri2020}
{Volonteri} M.,  et~al., 2020, arXiv e-prints, \href
  {https://ui.adsabs.harvard.edu/abs/2020arXiv200504902V} {p. arXiv:2005.04902}

\bibitem[\protect\citeauthoryear{{Wise}, {Regan}, {O'Shea}, {Norman}, {Downes}
  \& {Xu}}{{Wise} et~al.}{2019}]{Wise19}
{Wise} J.~H.,  {Regan} J.~A.,  {O'Shea} B.~W.,  {Norman} M.~L.,  {Downes}
  T.~P.,   {Xu} H.,  2019, arXiv e-prints, \href
  {http://adsabs.harvard.edu/abs/2019arXiv190107563W} {}

\bibitem[\protect\citeauthoryear{{Yang} et~al.,}{{Yang}
  et~al.}{2020}]{Yang2020}
{Yang} J.,  et~al., 2020, \mn@doi [\apjl] {10.3847/2041-8213/ab9c26}, \href
  {https://ui.adsabs.harvard.edu/abs/2020ApJ...897L..14Y} {897, L14}

\bibitem[\protect\citeauthoryear{{Yoshida}, {Omukai}  \& {Hernquist}}{{Yoshida}
  et~al.}{2008}]{Yoshida08}
{Yoshida} N.,  {Omukai} K.,   {Hernquist} L.,  2008, \mn@doi [Science]
  {10.1126/science.1160259}, \href
  {http://adsabs.harvard.edu/abs/2008Sci...321..669Y} {321, 669}

\bibitem[\protect\citeauthoryear{{de Bennassuti}, {Schneider}, {Valiante}  \&
  {Salvadori}}{{de Bennassuti} et~al.}{2014}]{deBennassuti14}
{de Bennassuti} M.,  {Schneider} R.,  {Valiante} R.,   {Salvadori} S.,  2014,
  \mn@doi [\mnras] {10.1093/mnras/stu1962}, \href
  {http://adsabs.harvard.edu/abs/2014MNRAS.445.3039D} {445, 3039}

\bibitem[\protect\citeauthoryear{{de Bennassuti}, {Salvadori}, {Schneider},
  {Valiante}  \& {Omukai}}{{de Bennassuti} et~al.}{2017}]{deBennassuti17}
{de Bennassuti} M.,  {Salvadori} S.,  {Schneider} R.,  {Valiante} R.,
  {Omukai} K.,  2017, \mn@doi [\mnras] {10.1093/mnras/stw2687}, \href
  {https://ui.adsabs.harvard.edu/abs/2017MNRAS.465..926D} {465, 926}

\makeatother
\end{thebibliography}


\bsp	
\label{lastpage}
\end{document}